\documentclass[journal]{IEEEtran}
\usepackage{savesym}
\savesymbol{checkmark}
\usepackage{dingbat}
\usepackage{rotating}
\usepackage{epstopdf}
\usepackage{color}
\usepackage{mathtools}
\usepackage{amsmath}
\usepackage{amsfonts}
\usepackage{amssymb}
\usepackage{enumerate}
\usepackage{amssymb} 
\usepackage{scrextend}
\usepackage{amssymb,amsmath,amsthm,enumitem}

\usepackage{pifont}
\newcommand{\xmark}{\ding{55}}%

\usepackage{multirow}

\usepackage{listings,inconsolata}
\usepackage{longtable}

\usepackage{comment}
\usepackage{verbatim}

\usepackage{tabularx,booktabs}
\usepackage{dingbat}
\usepackage{diagbox}
\newcolumntype{C}{>{\centering\arraybackslash}X}
\usepackage{caption} 

\usepackage{amsmath}
\usepackage{algorithm}
\usepackage[noend]{algpseudocode}

\usepackage{xcolor}

\usepackage{pdflscape}
\usepackage{afterpage}

\usepackage{amsfonts}
\usepackage{latexsym}
\usepackage{hyperref}
\usepackage{enumerate}
\usepackage{wasysym}
\usepackage{lipsum}
\usepackage{amssymb}
\usepackage{amsthm}

\ifCLASSINFOpdf
\begin{document}

\title{Blockchain for IoT Access Control: Recent Trends and Future Research Directions}

%

\author{Shantanu~Pal,~\IEEEmembership{Member IEEE},
		Ali~Dorri,~\IEEEmembership{Member IEEE},
	    and~Raja~Jurdak,~\IEEEmembership{Senior~Member,~IEEE}

\thanks{S. Pal, A. Dorri, and R. Jurdak are with the Trusted Networks Lab, School of Computer Science, Queensland University of Technology, Brisbane, QLD 4000, Australia (e-mail: shantanu.pal@qut.edu.au, ali.dorri@qut.edu.au, r.jurdak@qut.edu.au).}

}

\maketitle


\begin{abstract}
With the rapid development of wireless sensor networks, smart devices, and traditional information and communication technologies, there is tremendous growth in the use of Internet of Things (IoT) applications and services in our everyday life. IoT systems deal with high volumes of data. This data can be particularly sensitive, as it may include health, financial, location, and other highly personal information. Fine-grained security management in IoT demands effective access control. Several proposals discuss access control for the IoT, however, a limited focus is given to the emerging blockchain-based solutions for IoT access control. In this paper, we review the recent trends and critical needs for blockchain-based solutions for IoT access control. We identify several important aspects of blockchain, including decentralised control, secure storage and sharing
information in a trustless manner, for IoT access control including their benefits and limitations. Finally, we note some future research directions on how to converge blockchain in IoT access control efficiently and effectively.
\end{abstract}


%
\IEEEpeerreviewmaketitle

\section{Introduction}
\label{intro}
\begin{figure}[t]
 \centering
    \includegraphics[scale=.3]{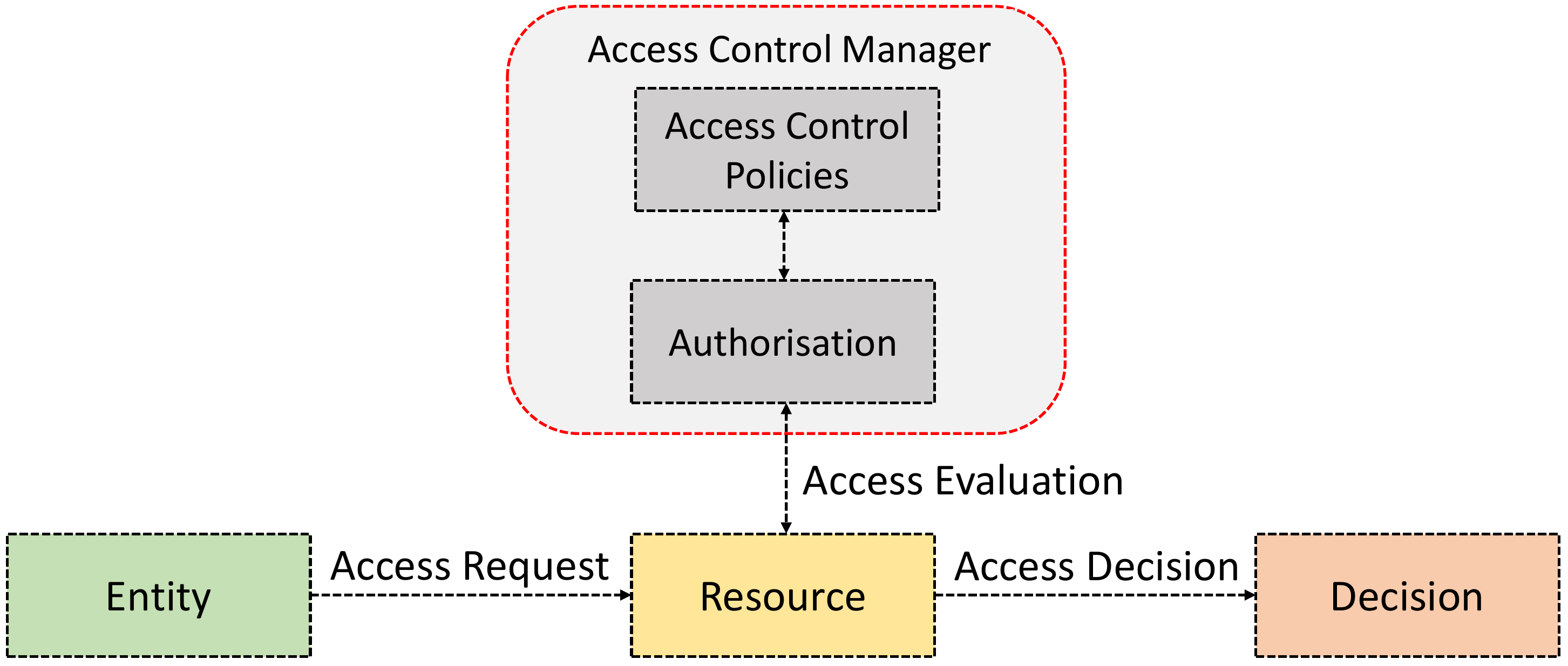}
    \caption{Major components of an access control process.}
    \label{fig:ac-1}
    \small
 \end{figure}

The scale of the number of devices, applications, users, and their associated services in an Internet of Things (IoT) system is massive~\cite{8286847}. It is predicted that there will be 50 billion smart devices connected to the Internet by the end of 2022. This will, in effect, increase the average number of devices and connections per household and Internet user. The annual global traffic is also predicted to reach 3.3ZB (Zettabyte) per year by the end of 2021~\cite{cisco-2021}. While the spread of various IoT applications provides better services, reduced cost of applications, and improved user experience, they pose significant security challenges to the system~\cite{lin:iot-security-privacy-survey-2017}~\cite{alqassem:iot-security-requirements-2014}~\cite{8086136}. In IoT, among other security issues, the question of access control is paramount~\cite{ANDALOUSSI20181031}. Access control can be seen as a security mechanism that ensures the reliable access of resources only by the authorised entities governed by a set of access control policies. It places a selective restriction of access that regulates who (e.g., an entity) can access or what (e.g., a resource) can be accessed under certain conditions.~\cite{tolone2005access}. In Figure~\ref{fig:ac-1}, we illustrate an outline of major components of an access control process. 

In IoT, the limited/portable device size, battery energy, and processing speed increase the device's vulnerability to network attacks. This increased vulnearbility stems from the inability to use  well-established conventional security mechanisms  directly to  resource-constrained IoT devices~\cite{dabbagh:iot-security-2017}. Additionally, the scale and heterogeneity of devices in IoT networks make it difficult to specify, centrally and in advance, a complete set of access control policies for both the users and devices. 

\subsection{Problem Description and Motivation}
\begin{figure*}[t]
 \centering
    \includegraphics[scale=.5]{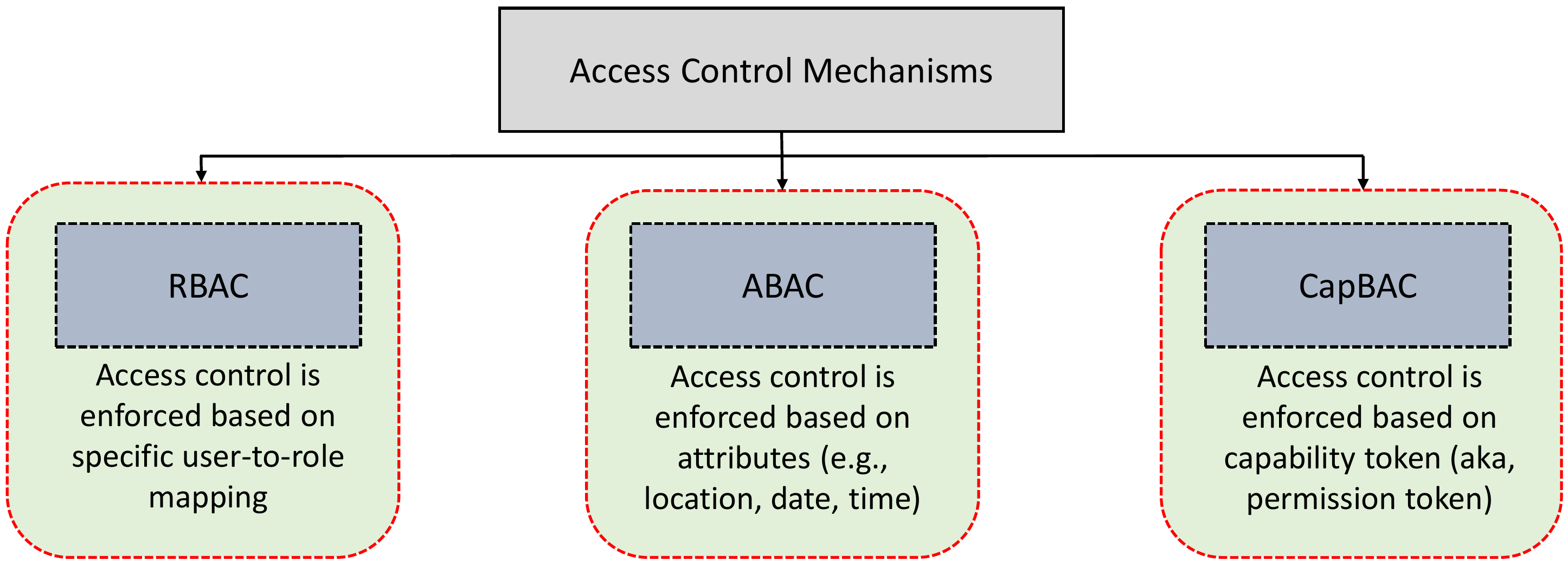}
    \caption{Commonly used access control mechanisms for the IoT.}
    \label{fig:ac-2}
    \small
 \end{figure*}

Most of today's access control mechanisms in IoT are developed on three commonly used access control mechanisms: Role Based Access Control (RBAC); Attribute Based Access Control (ABAC); and Capability Based Access Control (CapBAC). An outline of these three access control mechanisms are illustrated in Figure~\ref{fig:ac-2}. 

The employment of RBAC can provide a fine-grained access control over the resources explicitly using user-to-role mappings. However, in RBAC, for every access to a resource, there is a need to define separate user-to-permission relation. Moreover, RBAC is highly centralised in nature, limiting its \textit{scalability} for large-scale dynamic systems like the IoT. This further brings challenges towards the fine-grained \textit{permission enforcement} and \textit{attribute management} within the IoT systems to perform access control decisions. ABAC uses attributes to improve the policy management rather than depending upon the concrete unique identity of individual entities. This is promising given that the policies are written based on the context. In an IoT scenario this provides flexibility in policy management but at the same time ABAC does not provide any mechanism for controlling the number of policies required. 
In other words, ABAC does not, for example, support mechanisms for grouping together policies with different attribute requirements that allow an access to a single  resource or policies with the same attribute requirements that allow access to different resources. That said, ABAC requires a policy management mechanism for efficient \textit{resource management} and \textit{permission enforcement}, especially when the number of policies rises significantly. This creates an issue for scalable distribution of access control mechanisms to define a set of attributes uniquely acceptable for each user, device and service in a dynamic system like the IoT. CapBAC provides flexible access control by distributing capability tokens (also known as permission tokens) that contain access rights or privileges (along with additional conditions). These capability tokens can be validated at the edge IoT devices at the time of access to a resource. In this case, the edge IoT devices do not need to manage complex sets of policies. However, most of the CapBAC systems are centralised for policy storing and their management~\cite{pal2019policy}. This further creates challenges for the efficiency in \textit{attribute management} and flexibility in \textit{access rights transfer} at scale.

The discussion above gives rise to five key features of access control mechanisms in IoT: (1) resource management, (2) access rights transfer, (3) permission enforcement, (4) attribute management, and (5) scalability. These five features are critical to characterise access control, ranging from the management of resources to access control policy enforcement with proper security controls as well as delegating access rights from one entity to another in large-scale IoT systems. A detailed discussion of these categories is given in Section~\ref{blockchain-iot-papers}. Most of the current access control approaches so far address some but not all of these features adequately. 

\begin{figure*}[t]
 \centering
    \includegraphics[scale=.46]{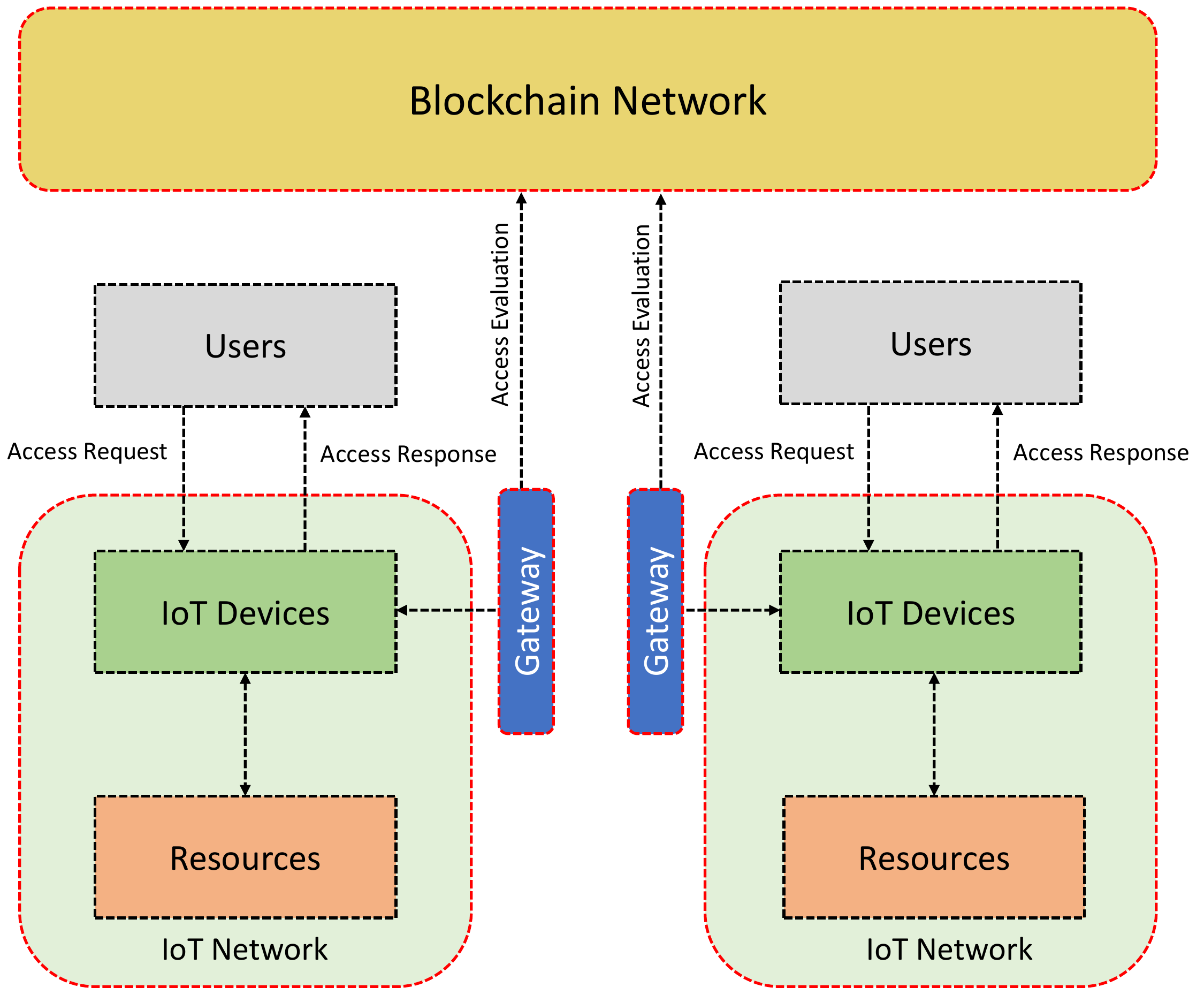}
    \caption{Access control in IoT based on blockchain technology.}
    \label{fig:ac-3}
    \small
 \end{figure*}

We note that access control in IoT requires consideration at design phase critical IoT requirements, so that it can provide scalable, efficient, light-weight, trustworthy and robust policy enforcement mechanisms. The large-scale and heterogeneity of IoT networks demands the decentralisation of policy management~\cite{MAJEED2021103007}. 
In recent years with the development of blockchain technology, it can be seen that the blockchain has the potential to address these issues in overcoming the limitations of traditional access control mechanisms in a more efficient and fine-grained way for large-scale IoT systems spanning multiple jurisdictions~\cite{wang2019survey}. In Figure~\ref{fig:ac-3}, we depict a conceptual view of blockchain-based IoT access control. Blockchain delivers new opportunities by providing distributed storage and a computational framework on which arbitrary programs can be executed. Several properties of the blockchain (e.g., no central authority and trusted third party, consensus mechanism, immutable, irreversible and tamper-proof, accessibility, auditability, etc.) offers a secure and safe way to record and track a list of transactions for a large number of devices in a highly transparent, auditable and efficient way by maintaining a peer-to-peer network. This cannot simply be achieved by the aforementioned commonly used access control mechanisms e.g., RBAC, ABAC, and CapBAC~\cite{101145-3180457-3180458}~\cite{101145-3350546-3352561}.

\subsection{Contributions}
Previous surveys in IoT access control (e.g., \cite{fotiou2016access},  \cite{elsayed2016access}, \cite{ranjan:iot-access-control-survey-2016}, \cite{zhang2016access}, \cite{8038503}, \cite{RAVIDAS201979}, \cite{Bertin2019}) mostly discussed the traditional view of access control mechanisms over a centralised infrastructure. These discussions are limited to the commonly used access control mechanisms e.g., RBAC, ABAC, and CapBAC. These surveys mainly point to access control aspects from the perspective of various security requirements e.g., confidentiality, integrity, and availability. However, they are more focused on general pervasive environments and limited focus is given to significant IoT-related issues e.g., context awareness, interoperability, transiency, scalability, trust, and decentralisation. 

While a few other proposals (e.g., \cite{ouaddah:iot-access-challenges-2017}, \cite{riabi-8766453}, \cite{343113734566789}, \cite{corr-abs-1908-08503}, \cite{sym12101663}, \cite{9223297}, \cite{8968396}) try to survey blockchain convergence in access control for the IoT, they do not focus on the various distinct properties of blockchain and their integration into IoT access control to a fine-grained level. Their discussions do not focus on key attributes that determine the suitability of an access control approach for a given context, e.g., how access control manages resources, whether it transfers access rights, how it enforces permissions or manages attributes, and whether it is scalable.

\begin{table*}[t]
\small
\centering
\caption{Previous surveys on IoT access control and their comparison with our work (BC = blockchain, AC = access control)}
\label{tab:comparison-previous-survey}
\begin{center}
\footnotesize
    \begin{tabular}{ p{1.5cm} p{1.5cm} p{1.7cm} p{1.7cm}  p{1.7cm} p{1.7cm} p{1.7cm} }
    \hline
    
    \multirow{2}{*}{Reference} &\multirow{2}{*}{BC-Based} & \multicolumn{5}{c}{BC-based solutions for IoT (functional classification)} \\ \cline{3-7}
    
    {} & {AC Issues} & \raggedright{Resource Management} & \raggedright{Access Rights Transfer} & \raggedright{Permission Enforcement} & \raggedright{Attribute Management} & Scalability \\ [0.5ex] \hline \hline
   
\cite{fotiou2016access} & \xmark & \xmark & \xmark & \xmark & \xmark & \xmark \\[0.5ex]  

\cite{elsayed2016access} & \xmark & \xmark & \xmark & \xmark & \xmark & \xmark \\[0.5ex]   

\cite{ranjan:iot-access-control-survey-2016} & \xmark & \xmark & \xmark & \xmark & \xmark & \xmark \\[0.5ex] 

\cite{zhang2016access} & \xmark & \xmark & \xmark & \xmark & \xmark & \xmark \\[0.5ex] 

\cite{8038503} & \xmark & \xmark & \xmark & \xmark & \xmark & \xmark \\[0.5ex]   

\cite{RAVIDAS201979} & \xmark & \xmark & \xmark & \xmark & \xmark & \xmark \\[0.5ex]  

\cite{Bertin2019} & \xmark & \xmark & \xmark & \xmark & \xmark & \xmark \\[0.5ex] 

\cite{ouaddah:iot-access-challenges-2017} & \checkmark & \xmark & \xmark & \xmark & \xmark & \xmark \\[0.5ex] 
    
\cite{riabi-8766453}  & \checkmark & \xmark & \xmark & \xmark & \xmark & \checkmark \\[0.5ex] 

\cite{343113734566789}  & \checkmark & \xmark & \xmark & \xmark & \xmark & \checkmark \\[0.5ex] 

\cite{corr-abs-1908-08503}  & \checkmark & \xmark & \xmark & \xmark & \xmark & \checkmark \\[0.5ex] 

\cite{sym12101663}  & \checkmark & \xmark & \xmark & \xmark & \xmark & \xmark \\[0.5ex] 

\cite{9223297}  & \checkmark & \xmark & \xmark & \xmark & \xmark & \xmark \\[0.5ex]

\cite{8968396} & \checkmark & \xmark & \xmark & \xmark & \xmark & \checkmark \\[0.5ex] 

\textbf{[Our work]} & \checkmark & \checkmark & \checkmark & \checkmark & \checkmark & \checkmark \\[0.5ex] \hline 
   
    \end{tabular}
\end{center}
\end{table*}

In this paper, we study the key features of blockchain technology (e.g., decentralisation, distributed ledgers, consensus, auditability, immutability, etc.) that makes it attractive for IoT access control and address several challenges (e.g., interoperability, inefficiency, and lack of trust) in the conventional mechanisms. In other words, our work takes the crucial features of blockchain technology to explore how they address the limitations of traditional access control mechanisms for large-scale IoT systems. Our work is intended to provide an outline on how to efficiently converge blockchain to improve IoT access control. We  focus on the key benefits and future research issues via this integration.

 In Table~\ref{tab:comparison-previous-survey}, we compare our present work to the previous surveys based on the two core categories: (a)  whether the proposals discuss the blockchain technology and how various properties of blockchain can satisfy the access control needs in IoT; and (b) the proposed blockchain-based access control solutions for the IoT. For the latter, we use the five key features of access control in IoT discussed above.  The major contributions of the paper can be summarised as follows:

\begin{itemize}

\item We provide a systematic literature review of the existing blockchain-based solutions for IoT access control. 
Our work bridges the gap between the access control need in distributed environments and the traditional highly-centralised mechanisms -- it highlights, investigates, and discusses the usefulness of blockchain in IoT access control.

\item Our work consider five key features of an access control mechanism that are significant to consider in blockchain-based access control solutions for the IoT. We provide a critical analysis of how these are satisfied in the existing literature. 

\item We provide a set of unique future research directions that help to efficiently integrate blockchain-based solutions in IoT access control.  

\end{itemize}

\subsection{Organisation and Roadmap}
The rest of the paper is organised as follows. In Section~\ref{access-iot}, we discuss the importance of access control for the IoT systems. This includes the discussion of specific access control needs in IoT.  In Section~\ref{access-control-for-iot}, we present the blockchain-based solutions for IoT access control. At first, we discuss blockchain technology in brief. Then we review the existing blockchain-based IoT access control solutions by categorizing the specific access control issue they address. In Section~\ref{open-questions}, we provide a discussion of lessons learned. We also highlight the future research directions for using blockchain in IoT access control. Finally, we conclude the paper in Section~\ref{conclusion}.

\section{Importance of access control in  IoT}
\label{access-iot}
With the rapid improvements in the IoT, there is a huge growth in the number of devices per user in recent years~\cite{8340813}. Moreover, the intelligence of smart IoT devices to sense, connect, and communicate with other devices make it more promising to apply in many areas than ever before. In such a context, anything and everything can be part of the network (e.g. via the Internet). This emphasizes the pervasive instrumentation of physical objects combined with smart devices~\cite{kevin-ashton:iot-2009}. In the IoT, devices and \textit{things} may, over their lifetime, interact with a vast range of other \textit{things}. Such interactions may be fleeting and may only occur once between a particular pair of \textit{things} or be much more frequent and long term~\cite{8489954}. \textit{Things} will likely be highly mobile, especially devices, moving from administrative domain to another administrative domain. These domains will have to establish policies and mechanisms to enable them to deal with devices and \textit{things} about which they have very limited, if any, previous information~\cite{8470752}. Note, for our purposes a \textit{thing} is one or a set of users, devices, services and applications, and similar entities.

IoT systems may deal with high volumes of data. This data can be particularly sensitive, as it may include health, location, and other personal information~\cite{8993839}. Activities that are currently not digitally enabled will be supported and others expanded by the edge intelligence and ubiquity of the devices that constitute the IoT~\cite{wang:iot-health}.  For example, shopping may be enhanced by services offered by \textit{things} deployed by the retailers, contacting user devices and offering information and discounts. Current services, e.g., e-tickets, may be enhanced by sensors detecting e-ticket holders and controlling physical access on that basis. Healthcare may be expanded by a range of sensors attached to a person~\cite{101145-31444573144485}.

Many of the vulnerabilities in IoT are associated with, (1) the identity of the \textit{things}, (2) trust management between the users and devices, (3) different network domains, and (4) dynamic network topology where the interactions between the \textit{things} may happen only once and or a very short interval of time. Even the unavailability of a network connection where devices are unable to get the latest software patches can generate attacks on these resource-limited devices~\cite{chahid:iot-security-2017}. There are several other characteristics of IoT e.g., openness, data freshness, self-healing etc. which increases the complexity and difficulty in protecting an IoT system from potential threats and attacks using conventional security mechanisms~\cite{8716500}. 

Attacks on IoT systems are fundamentally different from the traditional security and privacy related attacks in general computing systems. In IoT, attacks are becoming more sophisticated in terms of their mechanisms and the way they infect the system~\cite{8897627}~\cite{Patel2019}~\cite{mahler:know-your-enemy-iot-2018}~\cite{sun2018security}. This is not simply limited to penetrate a network layer with malicious codes or divert network traffic to another insecure destination without the knowledge of the users.  It is more pronounced where an IoT-enabled medical device can be compromised and controlled remotely by the attacker. For instance, a patient's pacemaker can be used to generate a fatal shock, or a drug infusion pump (e.g. insulin or antibiotics) can be controlled by an attacker to change the drug dosage with the authorized access~\cite{wired-iot-hack-2015}. In 2016, an attack called `Mirai Botnet'~\cite{mirai-2016} infected numerous IoT devices (in particular older routers and IP cameras) then flooded them with network traffic with a DDoS (Distributed Denial-of-Service) attack. In 2017, `Cayla' doll~\cite{doll-hack-2015} was banned in Germany for its immense privacy and security concerns. Cayla is an IoT-connected doll that provides children with a connected play experience by listening and talking to them. However, it can be a potential privacy threat as the dolls can be heavily compromised due to its insecure nature of Bluetooth connection. In Finland, in 2016, there was a complete shut down to the central heating and hot water systems using DDoS attack~\cite{findland-2016}. These incidents show the range of potential attack scenarios where a common IoT device can be compromised to infiltrate and attack larger networks.

 \begin{figure*}[t]
 \centering
    \includegraphics[scale=.75]{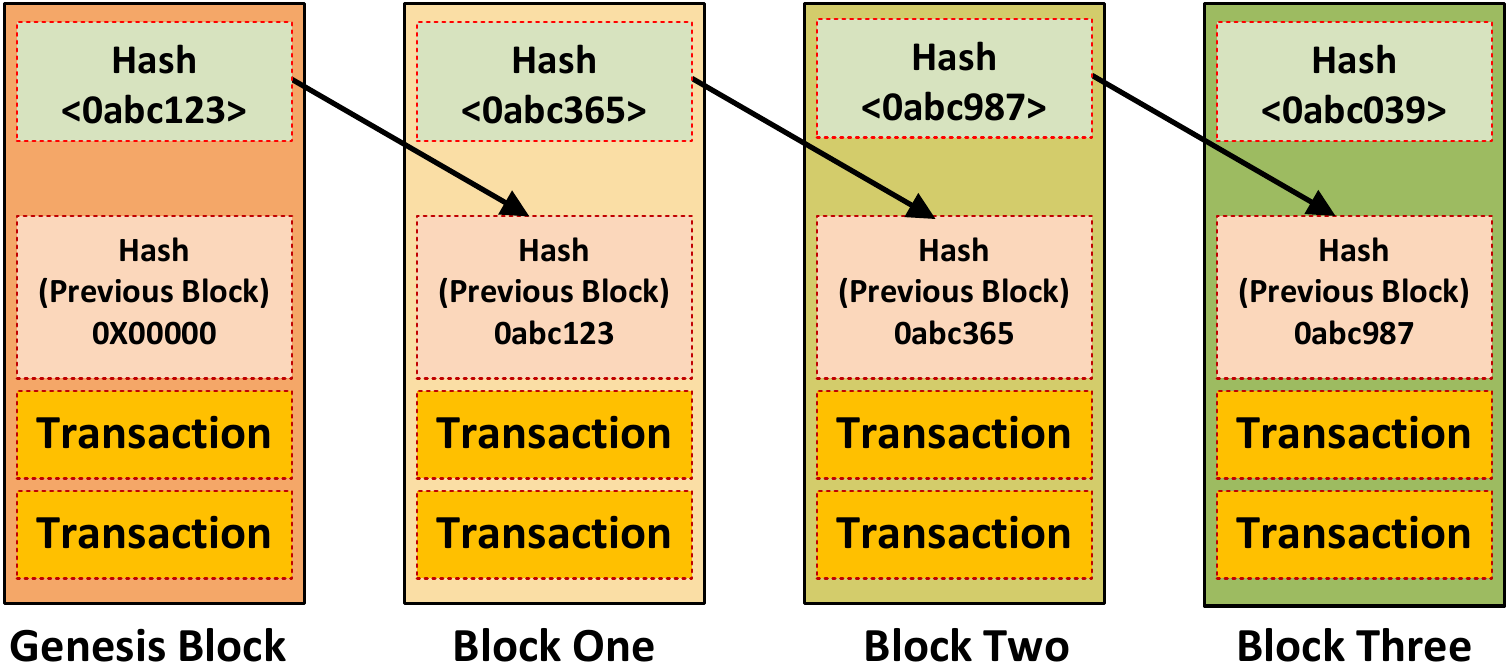}
    \caption{A simple view of formation of blocks in a blockchain network.}
    \label{blockchain-view}
    \small
 \end{figure*}

 Therefore, placing appropriate security mechanisms and enforcing proper access control policies for the IoT systems become important issues~\cite{uganya2021survey}. Moreover, access control must be placed in a way that can easily reach the edge of the IoT devices. There is a need to decentralize architecture that supports efficient control of accessing resources with the minimum number of policy requirements. That said, there is a significant need to prevent and control the unauthorised flow of information, and to develop appropriate security mechanisms for IoT access control to ensure proper security foundation for an IoT system~\cite{kouicem2018internet}~\cite{ezema2018open}~\cite{conti2018internet}~\cite{8058363}. 
 As discussed above, the existing access control solutions for the IoT do not meet the critical access control requirements, for example, decentralisation of control, scalability, and trust to a fine-grained level. Unfortunately, having a central point is common to a number of access control mechanisms in the existing proposals. What is even more common in such proposals in the literature is a single, particular means of policy evaluation. We note that a platform using blockchain can be used to efficiently handle such issues. The use of blockchain can constitute another step towards a robust distributed access control system that would overcome many of the challenges associated with centralisation, either in the form of a single entity calculating access control policies, or reliance on a single method for access rights.

\section{Blockchain-based solutions for IoT access control}
\label{access-control-for-iot}  
In this section, first, we briefly discuss blockchain technology and then we present the various blockchain-based access control solutions for IoT proposed in recent literature. 

\subsection{Blockchain Technology}
Blockchain originated as  the fundamental technology behind Bitcoin, the first cryptocurrency~\cite{8029379}. Blockchain establishes a trusted network over untrusted participants where transactions, i.e., communications between nodes, is verified by all participants. Blockchain eliminates the need for centralised controllers as all participating nodes jointly manage the network by storing and verifying new transactions and blocks.  Blockchain achieves distributed management as all nodes maintain the history of transactions in the form of chained blocks. Blockchain is immutable as each block maintains the hash of the previous block in the ledger, thus, any modification to the previously stored data will be detected (see Figure~\ref{blockchain-view}). The first block in the ledger is known as the \textit{genesis} block. The blocks  are organised by logical time stamps and synchronized among other member nodes within the network. Particular nodes in the network, known as miners or validators, collect new transactions and append them in the blockchain in the form of a block after following a \textit{consensus algorithm}. The latter ensures that the validator of the next block is chosen randomly which in turn ensures the blockchain security. The consensus algorithm normally involves solving a puzzle which demands resources which in turn protects the network against malicious nodes that may flood the network with fake blocks. 

The consensus algorithm ensures all nodes agree on the valid state of the ledger in two steps: (1) validator selection: this step is basically known as consensus algorithm in the literature and refers to selecting the validator of the next block, and (2) ledger agreement: due to the distributed nature of the blockchain multiple nodes may generate the same block simultaneously leading to creation of a fork in the network. In such cases, the blockchain relies on the concept of \textit{the longest ledger} to achieve consensus over the state of the ledger. Once a fork happens, the validators pick one  block randomly. Eventually, one ledger will end up with more blocks which will be considered as the valid block in the network~\cite{8123011}.

Blockchain technology can enrich  IoT by providing a platform for sharing information in a trustless manner due to its salient features including  immutability, auditability, and accountability. The history of the transactions is permanently stored in the blockchain and is stored by the participating nodes which in turn introduces a high-level of auditability. The distributed nature of the blockchain complements IoT  in various ways including reliability, security, accountability, and scalability~\cite{7945805}. 

In recent years, blockchain-based access control received tremendous attention due to the fundamental features of blockchain including auditability, distributed management, trust, and immutability.  For example, multiple blockchain based solutions for IoT have been developed including, Bosch XDK (Cross Domain Development Kit)~\cite{bosch} for collecting real-time cross-domain data, and  Hyundai Digital Asset Company (HDAC)~\cite{hdac} for quick authentication and data storage between IoT devices.

 \begin{figure*}[t]
 \centering
    \includegraphics[scale=.55]{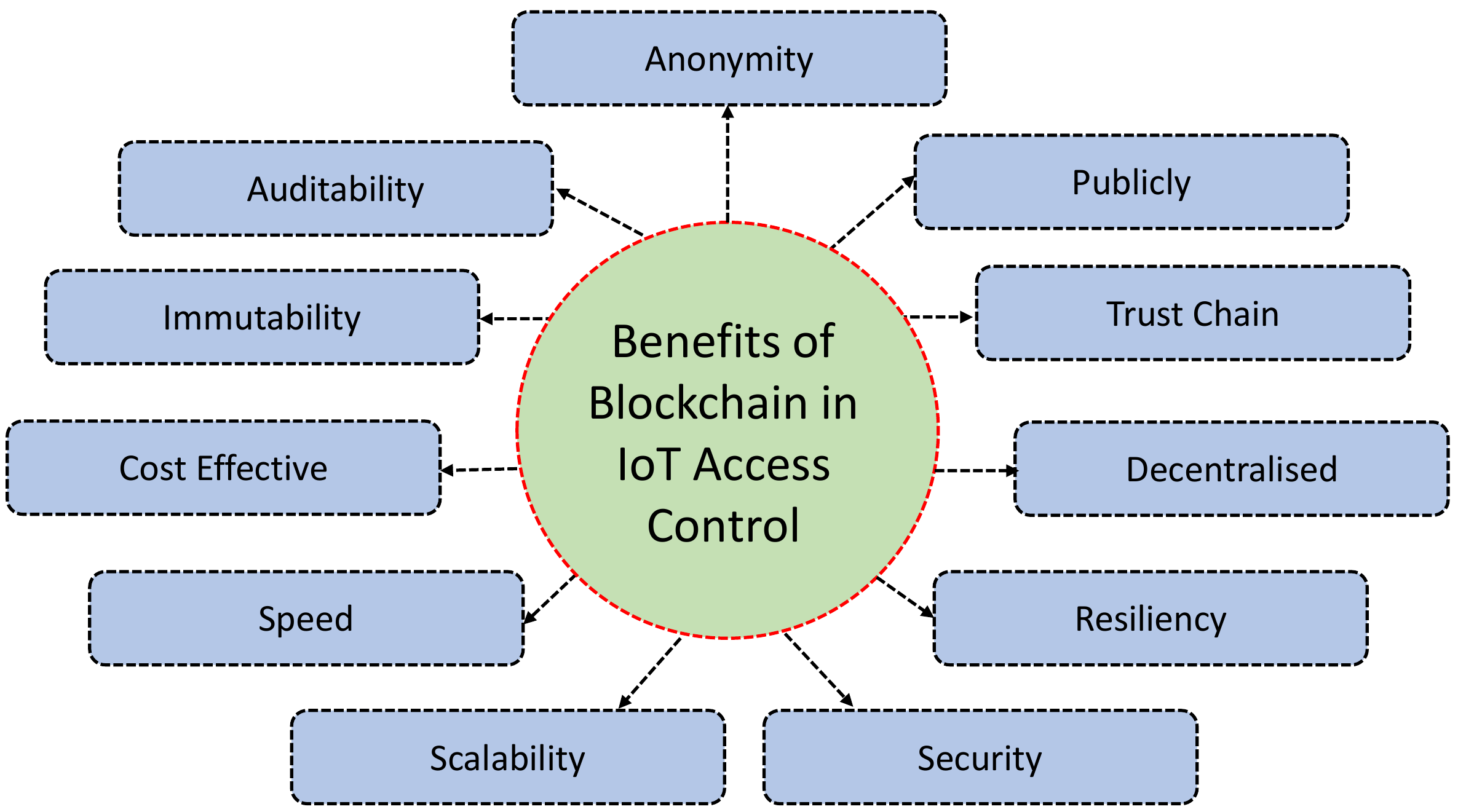}
    \caption{Several benefits of using blockchain for IoT access control.}
    \label{fig:ac-4}
    \small
 \end{figure*}

\subsection{Blockchain-Based Solutions}
\label{blockchain-iot-papers}
\begin{figure*}[t]
 \centering
    \includegraphics[scale=.5]{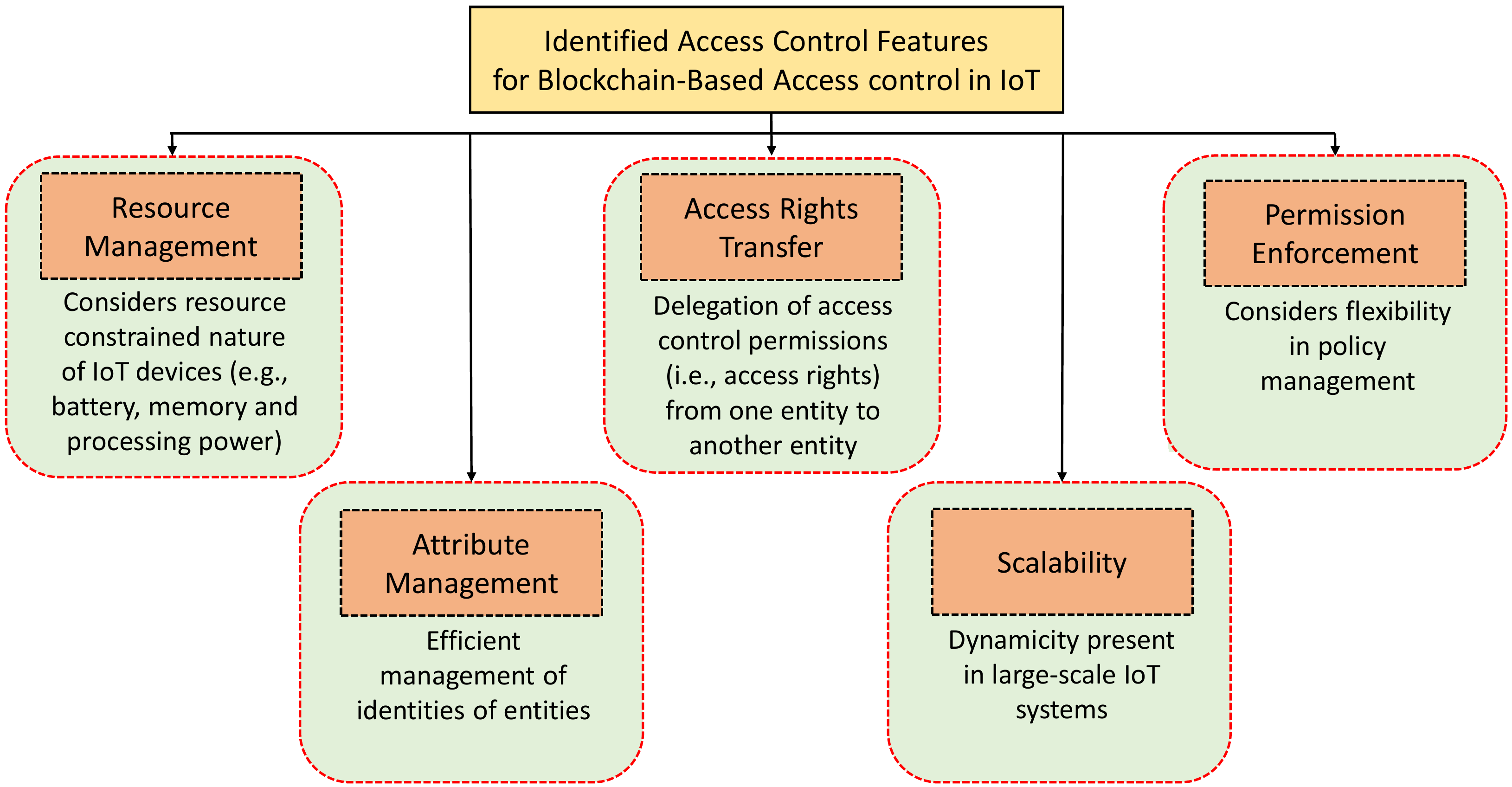}
    \caption{Critical features of access control identified for IoT in blockchain-based solutions.}
    \label{fig:ac-5}
    \small
 \end{figure*}

 The integration and use of blockchain with IoT is promising in several ways by ensuring the consistent ledger shared across the distributed network and verified transactions~\cite{8370027}~\cite{KHAN2018395}~\cite{DBLP-journals-07448}~\cite{s18082575}~\cite{REYNA2018173}~\cite{8580364}. This has advantages in managing access control mechanisms. In Figure~\ref{fig:ac-4}, we show various benefits blockchain in IoT access control~\cite{dedeoglu2020blockchain}~\cite{7917634}. In this section, we discussed the various blockchain-based access control solutions for the IoT across five critical features of access control that we find significant, they are, (1) \textit{resource management}, (2) \textit{access rights transfer}, (3) \textit{permission enforcement}, (4) \textit{attribute management}, and (5) \textit{scalability}. In Figure~\ref{fig:ac-5}, we show a summary of these features. Resource management is vital given the limited battery, memory, and processing capacity of IoT devices. Transfer of access rights denotes sending access control permissions (and any other conditions associated with that access) from one entity to another. It is important for enforcing access control delegation to the edge IoT devices. Permission enforcement must be tailored based on the needs of an IoT system. It is a key feature of access control systems to deal with flexibility in policy management. Attribute management, in particular, crucial given the uncertainty present an IoT system. The use of attributes can manage the identity of the entities (or even uncertainty in observations from physical and digital data) at scale that do not depend upon a unique concrete identity of an entity. Lastly, the importance of scalability in IoT access control cannot be overstated due to the robustness and dynamicity of data and resources, in particular, for the edge IoT/fog nodes. In Table~\ref{tab:table-solution}, we show the categorisation. %
A discussion of each of these issues along with the blockchain-based access control proposals are discussed as follows. 

\begin{table*}
\small
\centering
\caption{Comparison of various blockchain-based access control mechanisms proposed for the IoT based on the certain features of access control they addressed.}
\begin{tabular}{|c|c|c|c|c|c|}
\hline
{Reference} & {Resource} & {Access Rights} & \raggedright{Permission} & {Attribute} & {Scalability} \\ 
&Management & Transfer& Enforcement & Management&\\[0.5ex]\hline \hline 

\cite{8306880} & Yes & No & Edge & \raggedright{Smart contract } & No\\ 
&&&&(device identity)&\\ [0.5ex] \hline

\cite{8386853} & Yes & No & Edge & \raggedright{Smart contract} & No\\
&&&&(device identity)&\\[0.5ex] \hline

\cite{10-1007-978-3-319} & Yes & No & Centralised & \raggedright{Smart contract } & No \\ &&&&(device identity)&\\ [0.5ex] \hline

\cite{DBLP-1804-09267} & Yes & Yes (capability) & Edge & \raggedright{Smart contract } & No \\ &&&&(device identity)&\\ [0.5ex] \hline

\cite{8421332} & No & Yes (key-based) & Centralised & \raggedright{Smart contract } & No \\ &&&&(device identity)&\\ [0.5ex]\hline

\cite{8752021}~\cite{pal-iot-8894097} & Yes & Yes (capability) & Edge & \raggedright{Smart-contract } & Yes \\ &&&&(attribute-based)&\\ [0.5ex] \hline

\cite{ALI2019} & No & No & Edge & \raggedright{PoW} & No \\&&&&(trust-based)&\\ [0.5ex] \hline

\cite{Shafagh-2017-TBA} & No & No & Edge & \raggedright{Smart contract } & No \\ &&&&(enryption-based)&\\[0.5ex] \hline

\cite{algarni2021blockchain} & Yes & No & \raggedright{Edge} & \raggedright{Smart contract} & No \\&&&\& centralised &(device identity)&\\ [0.5ex] \hline

\cite{DORRI2019180} & No & No & \raggedright{Edge} & \raggedright{Smart contract} & Yes \\&&& &(device identity)&\\ [0.5ex] \hline

\cite{8964343} & No & No & Edge & \raggedright{Smart-contract } & Yes \\ &&&&(attribute-based)&\\[0.5ex] \hline

\cite{9355161} & No & No & Edge & \raggedright{Smart contract } & No \\ &&&&(attribute-based)&\\[0.5ex] \hline

\cite{zhang-electronics9020285} & No & No & \raggedright{Edge} & \raggedright{Smart contract} & Yes \\ &&&\& centralised&(attribute-based)&\\[0.5ex] \hline

\cite{almadhoun2018user} & No & No & Edge & \raggedright{Smart contract } & Yes \\ &&&&(device identity)&\\[0.5ex] \hline

\cite{DBLP-journals-048309} & No & No & Edge & \raggedright{Smart contract } & Yes \\ &&&&(device identity)&\\[0.5ex] \hline

\cite{8308029} & No & No & \raggedright{ Edge} & \raggedright{Smart contract } & Yes \\ &&&\& centralised&(device identity)&\\ [0.5ex]

\hline
\end{tabular}
\label{tab:table-solution}   
\end{table*}


\subsubsection{Resource Management}
\begin{figure*}[t]
 \centering
    \includegraphics[scale=.55]{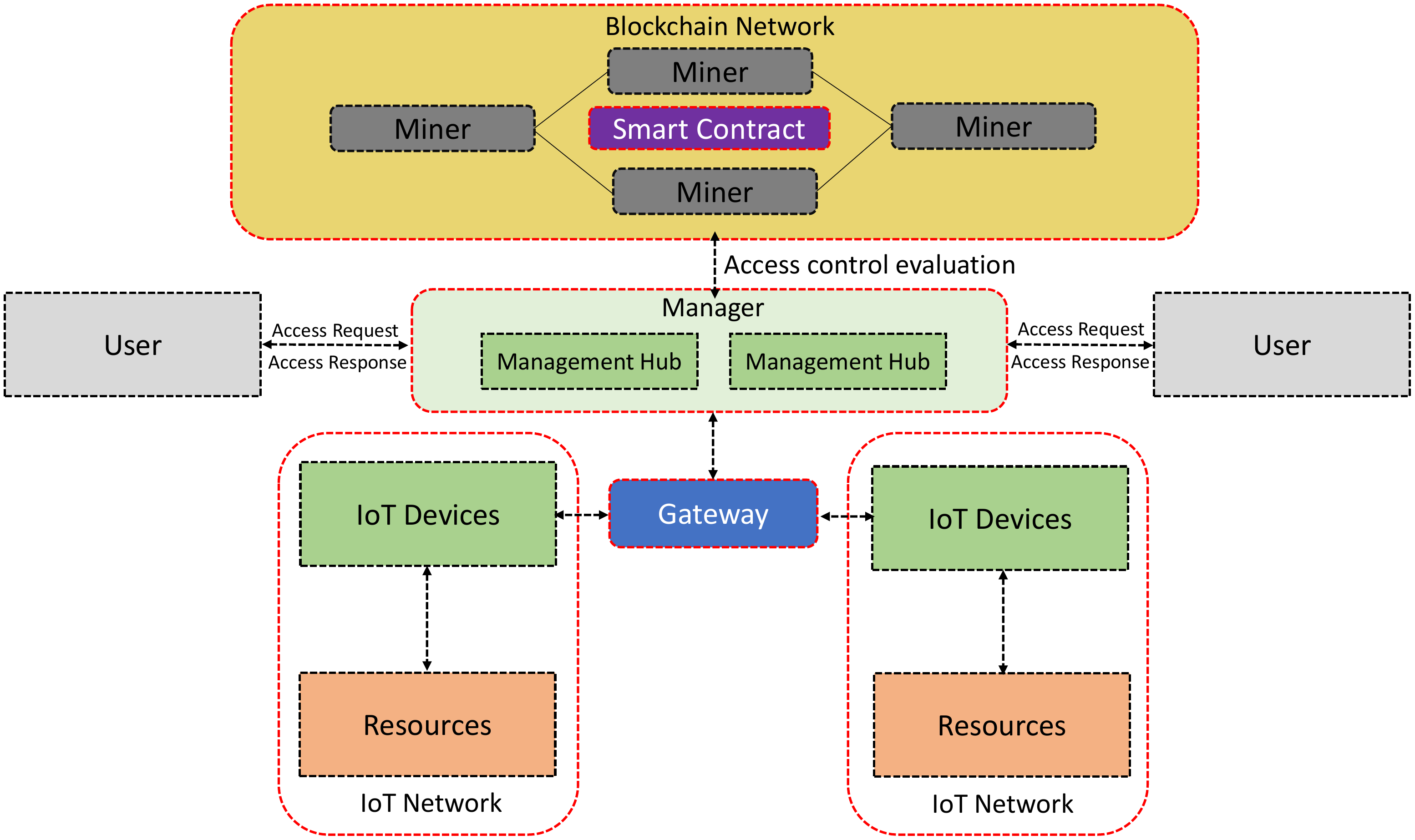}
    \caption{An IoT access control architecture based on blockchain presented in~\cite{8306880}.}
    \label{fig:ac-6}
    \small
 \end{figure*}

In an IoT access control perspective, smart devices will come in many forms and provide a vast range of services to both their users and other entities within the system. IoT systems should provision efficient management of computing storage/resources and their allocation and sharing. Users also wish to efficiently access these resources, and quickly and precisely obtain the services they require in a secure and distributed environment. The devices will, for the foreseeable future, present relatively low-power capacities and resource management requirements must be tailored to this~\cite{hameed2019understanding}. To this end, there is a high demand to build an efficient resource management framework that composed of a vast number of IoT devices. In order to provide a more flexible and adaptive resource management framework for access control in IoT, the use of blockchain can be an alternative to the commonly used centralised access control systems. Recall, that blockchain removes the control from a centralised node and provides more flexibility in resource management for a number of scenarios including supply chain, transportation, and energy sectors, consists of a vast amount of IoT devices. The use of blockchain in such cases cryptographically guarantees the data’s irreversible and unforgeable characteristics through smart contracts. To address the resource management issues in access control for large-scale IoT systems, Novo~\cite{8306880} presents an architecture for resource management of IoT devices supported by blockchain. The architecture is fully distributed in nature leverage the properties of blockchain technology (cf.~Figure~\ref{fig:ac-6}). The access control policies are enforced within the blockchain. The proposed design operates a single smart contract which reduces the communication overhead among the nodes, and at the same time significantly optimize resources. It also provides access control in real-time to the edge IoT devices. Note, all entities in the system are part of the blockchain network except the IoT devices. This is due to the resource-constrained nature of the IoT devices, where the devices cannot store the heavy-weight blockchain information. The proposed architecture is able to manage a vast amount of IoT devices and provide a decentralised feature of access control that connects a high number of geographically distributed sensor networks. The access control policies are enforced based on blockchain technology overcoming the bottleneck of a single centralised authority that manages the access control decisions. 

In this model, the edge IoT devices do not belong to the blockchain network, they are connected to the blockchain using one or more management hubs. These hubs are distributed over the entire blockchain network and potentially connected in different ways to the IoT devices which notably provide considerable flexibility in the overall access control by the utilisation of resources to a fine-grained process of access to data. Significantly, this model brings computing resources to the edge of the IoT network with a secure distribution of resources for the edge nodes. The use of blockchain simplify the network traffic in the core network, as well as provides a minimal end-to-end latency, response time and higher throughput. 

Another proposal~\cite{8386853} discusses a smart contract-based access control framework for IoT. In this framework, several `access control contracts' for access control between users and resources are computed inside the blockchain network. To control access between multiple resources, the access control contract validates the dynamic access rights based upon the behaviour of the subjects. The proposal discusses a resource sharing mechanism that takes the advantages of blockchain smart contracts. The user simply needs to store the access control rules for a given resource and the blockchain will manage access to that resource to the other users. It uses Representational State Transfer (REST) design pattern in combination with IoT Constrained Application Protocol (CoAP) to enable resource cooperation between the users. Similar to~\cite{8306880}, in this proposal access control management is performed within the blockchain network. 
Nuss et~al.~\cite{10-1007-978-3-319} present a blockchain-based storage and access management framework for large-scale IoT systems. This proposal first investigates the current challenges of identity, storage, and access management issues in IoT and then employs blockchain to examine how those challenges can be controlled. The proposal addresses the increased demand for secure and comprehensive storage and resource management issues in IoT as well as discusses the question of interoperability between heterogeneous and resource-constrained devices using blockchain technology. 
It uses block size adaptation scheme to address the resource allocation issue that operates in a distributed way to relieve the load of edge nodes. The proposed system enables efficient interoperability between the users and the devices by a variety of resource optimisation scheme that helps to identify the data sources that in turn assist reliable and efficiency of edge resource management. Further, it addresses the scalability issue in terms of network and storage consumption. 

\subsubsection{Access Rights Transfer}
Access control transfer  is significant as it helps to provide certain access from one entity to another with specific access rights. For instance, a mechanic being granted rights to a car’s systems is able to carry out maintenance as directed by the car’s owner. Generally, the transfer of access rights from one entity to another is known as delegation. In a delegation, the entity that transfers the access rights is known as the \textit{delegator} and the entity that receives the delegated access rights is known as the \textit{delegatee}. In large-scale and highly mobile systems, e.g., the IoT, delegation plays a vital role in ensuring flexible, fine-grained, and responsive access to resources by allowing users to propagate access in a controlled fashion. However, in the case of an IoT system it difficult to specify, centrally and in advance, the complete set of access control policies in a trustworthy manner. There are a few proposals that devise delegation of access rights in general IoT, but they overlook the crucial aspects of ownership of delegation. Recent proposals try to address this issue using blockchain where the propagation of delegation (especially, the delegation chain) can be easily verified. 

Proposal~\cite{DBLP-1804-09267} discusses an access control model for IoT using CapBAC to the resources, namely `BlendCAC'. In this access control model, a capability-based delegation mechanism is discussed for the propagation of access control permissions based on the blockchain network. In particular, the authorisation mechanism of delegation is computed inside the blockchain. 
Le and Mutka~\cite{8421332} propose a blockchain-based decentralised model for delegation access rights in IoT, named `CapChain'. This allows users to share and delegate their access rights efficiently and seamlessly to other IoT devices in public but still maintain privacy and user's identity by the secure distribution of keys leveraging the use of blockchain transactions. Similar to~\cite{DBLP-1804-09267}, this scheme uses capabilities for access rights delegation over the blockchain networks. Here, every IoT device in the network contains at least one owner who has full control over the device and is capable of generating capabilities based on the access control policies specified by the system. The capabilities are then transferred from one device to another via blockchain transactions. An experimental setup is provided with evaluation results to support the design. Unlike~\cite{DBLP-1804-09267}, which uses an identity-based capability token management strategy to protect users' privacy, proposal~\cite{8421332} uses an anonymous crypto-currency blockchain systems that ensures user’s privacy by hiding sensitive information (e.g., identity). 

\begin{figure*}[t]
 \centering
    \includegraphics[scale=.55]{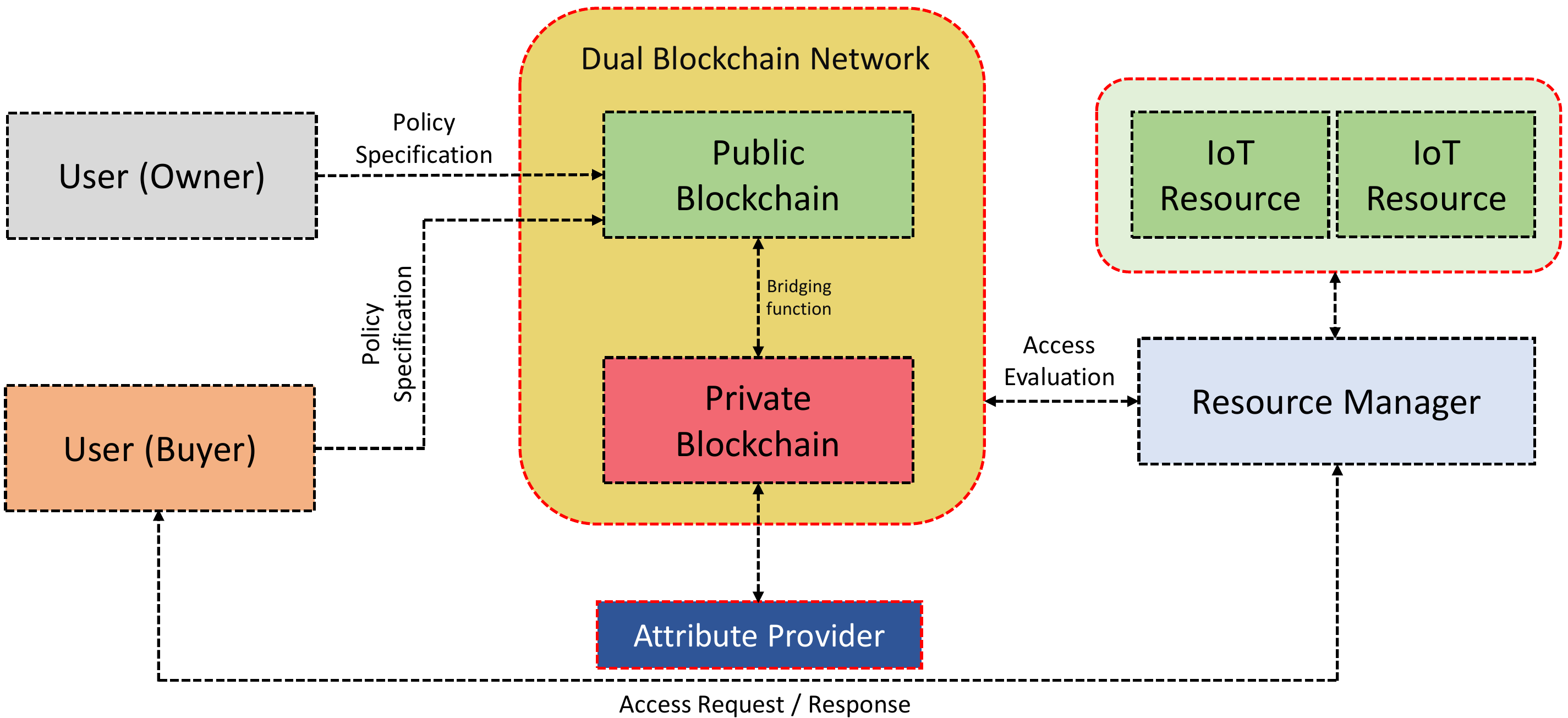}
    \caption{An IoT access control architecture based on a dual-blockchain concept presented in~\cite{8752021}.}
    \label{fig:ac-7}
    \small
 \end{figure*}

Pal et~al.~\cite{8752021} discuss the critical issue of access control delegation in an IoT system supported by blockchain technology and propose a flexible decentralised delegation model for transferring access rights in an IoT system. Using capability, access rights are transferred from one entity to another based on certain policies. Access rights (and other conditions) are embedded inside the capability token and the access rights (and the conditions) are evaluated at the edge IoT devices during the time of accessing a resource. In this proposal, an evaluation of access rights is performed outside the blockchain network. This is significant when we consider a large-scale system like the IoT and their associated policy settings for individual access control issues for edge IoT devices. In~\cite{pal-iot-8894097}, Pal et~al. extend the proposed blockchain-based access control platform of~\cite{8752021} to a `dual-blockchain' platform that combines a private and a public blockchain (cf. Figure~\ref{fig:ac-7}). This increases flexibility to protect users' privacy as the sensitive attributes are stored inside an attribute provider which is managed by the private blockchain. In other words, the proposed dual-blockchain architecture moves the attribute storage and access of the public blockchain onto a secure private blockchain. A bridging program is used based on a load balancing or leader election algorithm that links these two blockchains. Notably, these two blockchains maintain sustainable access control decisions independently. They provide adequate security and at the same time maintain the confidentiality and integrity of data. Both the proposals~\cite{8752021} and~\cite{pal-iot-8894097} are asynchronous, distributed, and use attribute-based identity when delegating access rights from one entity to another. In other words, the delegation is identity-less in nature, which argues for the use of non-unique identities in the delegation. This is important given the diverse nature and scale of an IoT system. Significantly, different from~\cite{DBLP-1804-09267} and~\cite{8421332}, in these cases the blockchain events are used as capabilities (also referred to as permission or access token) to facilitate access control delegation to IoT devices whereby the generated capabilities are issued by the smart contracts without the involvement of any trusted third-party authentication. An Internet business model (involving owner and buyer) is used to discuss the feasibility of the solutions in a real-world scenario. 
A detailed implementation is provided.

\subsubsection{Permission Enforcement}
Enforcing proper access control permissions among a large number of distributed IoT devices requires more coordination on the edge networks. The use of blockchain in such cases provides more flexibility and robustness.  Smart contracts are used to monitor and enforce of access control permissions under complex conditions~\cite{unal2020policy}. Furthermore, blockchain shows the potential of enforcing distributed access control permissions expressing the right to access a resource at a fine-grained level.
Ali et~al.~\cite{ALI2019} discuss an approach of permission-based access control for blockchain-based IoT systems. The proposed approach leverages the decentralised nature of blockchain for permission enforcement between the entities within the system. The main motivation of this study is to establish trust between the entities while removing the central trusted authority to maintain the trust degrees for each entity that controls and monitors access control enforcement. This is achieved by the use of PoW - the consensus mechanism of blockchain. To provide a light-weight solution, access rights are assigned to a node with a minimum number of permissions. No proof of concept implementation is detailed. 

Shafagh et al.~\cite{Shafagh-2017-TBA} present a  blockchain-based design for the IoT systems that provides distributed access control and access rights. The authors identify three requirements that are essential for such design: secure data storage, IoT compatibility, and decentralised access rights management. The design enhances blockchain technology to manage ownership and data sharing between the owners and the IoT devices. Owners can create new transactions to the blockchain that contains the identifier of the data stream and the service's public key. When a user wants to revoke access rights from a specific user, it changes the encryption key and shares the new key with all authorised services over the blockchain network, except the one that is revoked. This ensures flexibility in access control permission enforcement for large-scale IoT systems by introducing a set of separate permissions at each stage of an access control process. No implementation detail of the proposed design is provided.

Like~\cite{Shafagh-2017-TBA}, Algarni et~al.~\cite{algarni2021blockchain} present an access control model for the IoT based on blockchain. This provides a light-weight and decentralised secure access control framework for enforcing access control permissions using smart contracts. The core objective of the proposed model is to provide secure communication and trustworthy policy enforcement between the edge IoT devices supported by the underlying blockchain properties e.g., scalability, auditability, and transparency. A private hierarchical blockchain structure is taken into consideration to achieve more fine-grained permission enforcement in different level of access (e.g., at user level and blockchain level). In user level, cryptographic operations are used. In blockchain level, light-weight consensus mechanisms are used for permission enforcement based on various IoT requirements for access control.

Ali et~al.~\cite{DORRI2019180} propose a `Light-weight Scalable Blockchain' solution for low-resourced IoT devices known as LSB. It is employed to optimize IoT requirements and efficient permission enforcement using a Distributed Throughput Management (DTM) algorithm. LSB incorporates three levels of optimisations which are: (1) lightweight consensus algorithm that requires the validators to wait for a random period of time before committing new blocks to the blockchain, (2) distributed trust where the processing overhead required for verifying the transactions and blocks is reduced as nodes built trust on each other. The trust is measured based on the historical behavior of the nodes, and (3) a distributed throughput management algorithm that ensures blockchain can achieve self-scaling feature. LSB divides the participating nodes into clusters and the Cluster Heads (CH) manage the blockchain. LSB incorporates access control in blockchain-level where CH manage access to the cluster members. Each CH maintains an access control list that indicates which public keys (i.e., users in the overlay) can access a particular cluster member. The access control list is populated by the cluster members.

 \begin{figure*}[t]
 \centering
    \includegraphics[scale=.55]{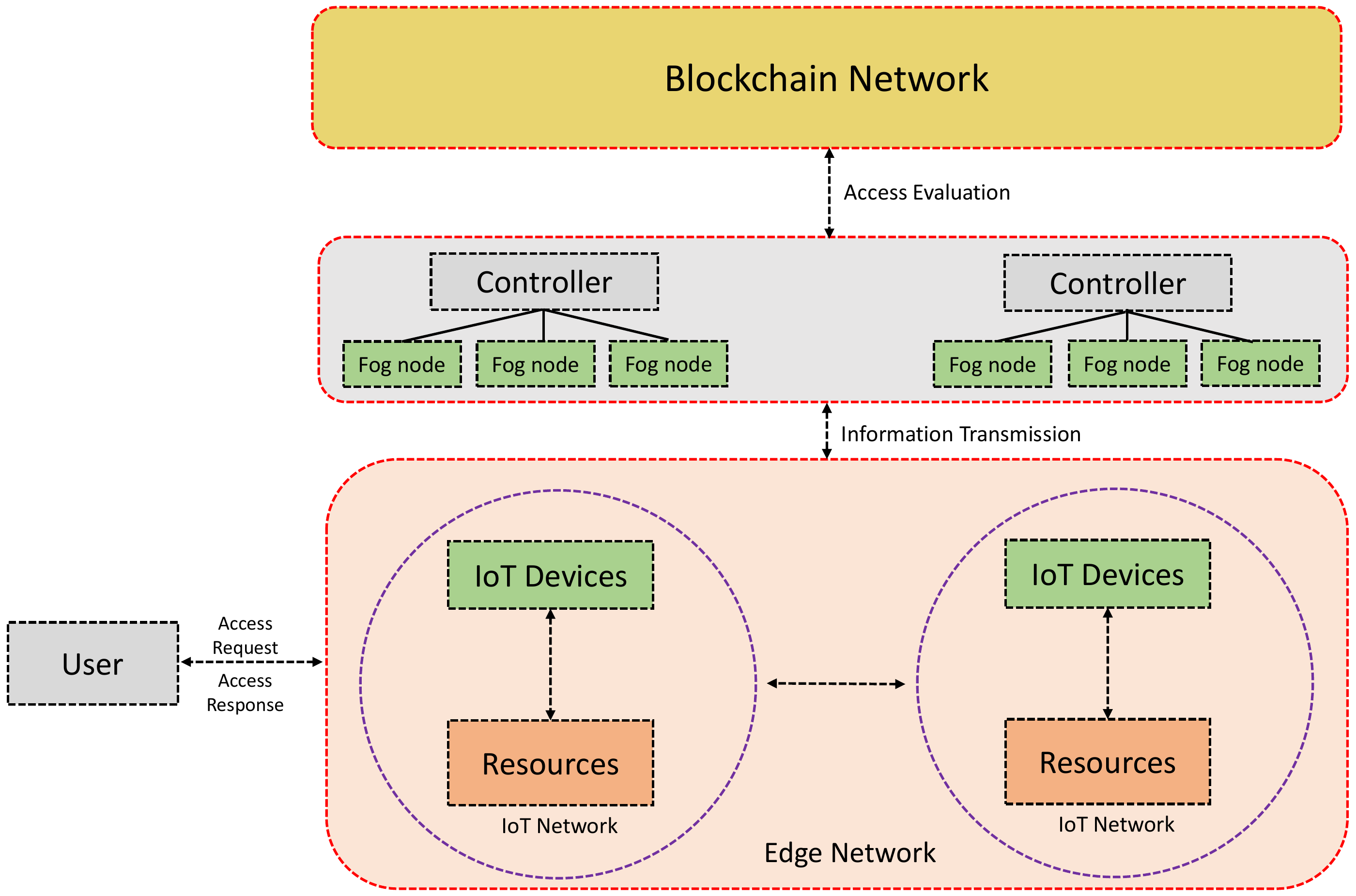}
    \caption{Fog computing for IoT access control in blockchain-based solutions.}
    \label{fig:ac-8}
    \small
 \end{figure*}

\subsubsection{Attribute Management}
Management of attributes (e.g., location, date, time, etc.) is significant to provide a decentralized, flexible, and fine-grained authorisation for IoT devices. Attributes are significant as they are used to express specified access policies by a target to decide if the requested entity fulfills the required privileges that are necessary for access.  Blockchain is utilized in such cases that allow authentic and reliable credentials. It is possible to use the blockchain property where verifiable collaboration mechanisms are controlled using edge IoT devices. Furthermore, blockchain is used to record and distribute attributes in order to avoid data tampering and a single point of failure.
Liu et~al.~\cite{8964343} present a decentralised, fine-grained, and dynamic access control system to provide efficient attribute management for large-scale IoT systems supported by blockchain. The proposed system is called \textit{`Fabric-IoT'}, which is based on the  Hyperledger Fabric blockchain framework and takes advantage of an ABAC system. In this model, smart contracts are created to control and manage the access control policies both for the admin and end-users. These three smart contacts are: (1) Device Contract (DC) (2) Policy Contract (PC), and (3) Access Contract (AC). DC provides a method that helps to store the unique URLs of selected resources generated by certain IoT devices. PC is responsible for managing and enforcing ABAC policies for admin users. Finally, AC contains the core access control methods that are used for the end-users. Simulation-based experimental studies are performed to show the applicability and to ensure data consistency of the proposed system in real-world IoT settings. 

With a similar concept of~\cite{8964343}, Sun et~al.~\cite{9355161} propose an attribute management framework for IoT access control supported by ABAC. In this proposal, the IoT system is divided into different functional domains. Then a local blockchain ledger has been established for each of the domains. The local blockchain ledger records domain entities, e.g., attributes, access decisions, etc. This will help to enable more IoT devices as blockchain nodes. It uses an identity-based signature to serve cross-domain access requests that are coming from legitimate users from each IoT domain.

Employing blockchain to implement an access control mechanism based on attributes, e.g., the one associated with ABAC, would likely require attributes to be stored in the blockchain. Storing of attributes to the blockchain raises questions of adequate privacy as all users can see all entries in the blockchain. To address this issue, Zhang et~al.~\cite{zhang-electronics9020285} propose an attribute-based access control framework that addresses the privacy-aware and efficient distribution and management of attributes using blockchain for large-scale IoT systems. Blockchain is used to provide a trustworthy environment for transmitting access information. In other words, blockchain is used to deliver authentic and reliable credentials. To support a controlled access authorisation a trusted collaborative mechanism is implemented. The collaboration has happened between the ABAC and blockchain. This proposal also provides a light-weight, decentralised access control solution where the IoT devices only need to store a string of access information to perform an authorisation task. 
It is also worth mentioning that the proposals~\cite{8752021} and~\cite{pal-iot-8894097} use attributes to validate an entity in the systems rather than depending upon their unique concrete identity. It is useful in many cases where an entity does not need to share or delegate their unique identity and can simply use the notion on attribute-based identity management.

\subsubsection{Scalability}
There is a growing trend that examines the suitability of blockchain technology with the emerging fog computing technologies for IoT access control~\cite{Kiwelekar2021}. In Figure~\ref{fig:ac-8}, we illustrate a conceptual view of employing fog computing for IoT access control in blockchain-based solutions. Recall that blockchain provides a distributed ledger for strong access to information and is decentralised in nature. This reinforces the decentralised requirement of access control in the IoT. The use of fog computing in IoT systems enhanced the use of traditional cloud computing technology for aggregating, processing, and analyzing heavy network traffic and workload. Fog nodes are used as a local controller located in a close distance of the edge IoT devices, acting as an intermediate layer between the edge IoT nodes and cloud servers~\cite{8314121}~\cite{Puliafito-Fog-2019}~\cite{10156084}~\cite{7033259}. Fog computing can be seen as an extension of cloud computing which provides a highly scalable and dynamic visualised platform for processing, strong and networking services between the cloud and the edge network~\cite{Bonomi-2342513}. Traditional cloud computing technology is widely used for aggregating, processing, and analyzing heavy network traffic and workload. However, in the case of IoT devices, cloud computing is not preferred platform in terms of responsiveness and intermediate processing of the IoT data. This further reinforces the resource-constrained nature of the IoT devices. In such a case, fog nodes can be seen as a local controller that is placed in close proximity to the edge IoT devices (e.g., geo-distributed fog servers) and it is primarily responsible for the local aggregation, processing, and analyzing the IoT data. As such, the fog architecture provides an intermediate layer between the edge IoT nodes and cloud servers. Moreover, 
fog provides efficient, delay-sensitive, and location-aware services for the edge IoT devices~\cite{9139976}.

Proposal~\cite{almadhoun2018user} discusses the concept of blockchain-enabled fog nodes that are used with the IoT devices to provide more flexible access control. This leverages flexibility to access control by considering the resource-constrained nature of the IoT devices. The fog nodes are used to enhance the scalability of the system where the heavy computational takes related to authentication and communicating with the blockchain network are carried out by the fog nodes. This improves the performance of the IoT devices. Fundamentally, the fog nodes interface to Ethereum smart contracts to authenticate the legitimate users to access resources (i.e., IoT devices). The proposed architecture is able to manage a vast amount of IoT devices and provide a decentralised feature of access control that connects a high number of IoT devices.  The access control policies are enforced based on blockchain technology overcoming the bottleneck of a single centralised authority that manages the access control decisions. In this model, the edge IoT devices do not connect to the blockchain network directly, instead, they are connected to the blockchain using one or more management hubs. These hubs are distributed over the blockchain and potentially connected in different ways to the IoT devices which significantly provide considerable flexibility in the overall access control management. 

In a similar vision to~\cite{almadhoun2018user}, Farhadi et~al.~\cite{DBLP-journals-048309} discuss a blockchain-enabled fog architecture to allow secure access control in IoT applications. For a set of for nodes a dedicated controller is defined. The proposed solution is based on a multi-data owner and multi-service provider environment. Where these are distributed over a wide area. A high conceptual view of a fog blockchain manager is discussed that is composed of fog nodes and a blockchain network govern by a set of controllers. A fog-to-fog consensus mechanisms is discussed to employ these security features in access control. No implementation is provided. In this work, IoT access control security is outlined with the following five dimensions: confidentiality, integrity, authenticity, non-repudiation, and availability. 

Riabi et~al.~\cite{8308029} discuss an approach for distributed access control over fog computing for the IoT systems. In this approach, IoT nodes are connected to a local controller (a fog controller) which is able to perform heavy computation processing and synchronizes with the cloud servers. These controllers can be seen as the collaborative agents that helps integration of IoT nodes and cloud platforms. This provides flexibility in access control of data to a more granular level that are coming from multiple sources. A unified set of access control policies dedicated to an access is supplied to each controller. This in turn reduces both administrative and processing overheads of the system by a scalable distribution of computational power and storage capacity.

\begin{table*}
\small
\centering
\caption{Summary of access control mechanisms for the IoT based on blockchain technology}
\begin{tabular}{p{1.2cm} p{.7cm} p{6.2cm} p{6.2cm} p{1.7cm}}
\hline   
    {Reference} & {Year} & {Purposes} & {Key Contribution} & {Implementation} \\ [0.5ex]\hline \hline

\cite{8306880} & 2018 & Building an access control model for IoT using blockchain considering the resource constrained nature of the IoT devices. & Develops an access control architecture for scalable management of IoT devices using blockchain. & Yes \\ [0.5ex] 

\cite{8386853} & 2019 & Examining the employment of smart contracts for IoT access control. & Proposes a smart contract based access control framework for IoT using supported by blockchain. & No \\ [0.5ex] 

\cite{10-1007-978-3-319} & 2018 & Investigate the storage, identity, and access management issues for large-scale IoT systems using blockchain-based solutions. &  Proposes a secure access control management framework for heterogeneous IoT systems supported by blockchain, especially focused on resource constrained IoT devices. &  No \\ [0.5ex]

\cite{DBLP-1804-09267} & 2018 & Examining access right delegation issue in IoT using blockchain. & Develops a delegation architecture called \textit{`BlendCAC'} for IoT using capabilities (as access tokens) and blockchain networks. & Yes \\ [0.5ex] 

\cite{8421332} & 2018 & Examining the use of capabilities (as access tokens) for transferring of access rights over the blockchain networks. & Develops an access control architecture called \textit{`CapChain'}. that allows users to share and delegate their access rights efficiently and seamlessly to IoT devices.  & Yes \\ [0.5ex]

\cite{8752021}~\cite{pal-iot-8894097} & 2020 & Examining the access control delegation issues (i.e., transferring of access rights from one entity to another) for large-scale IoT systems using blockchain. & Develops a duel-blockchain based access control architecture that is asynchronous in nature, decentralised and supported by attribute based identity. & Yes \\ [0.5ex]

\cite{ALI2019} & 2019 & Addressing the identity and access management issues in IoT using blockchain. & Presents a blockchain-based identity and access management framework for large-scale IoT systems. & No \\ [0.5ex]

\cite{Shafagh-2017-TBA} & 2017 & Examining the suitability of blockchain for distributed access control and data management in the IoT. & Proposes a decentralised access control management for information sharing in IoT using blockchain. & No \\ [0.5ex]


\cite{algarni2021blockchain} & 2021 & Building a secure communication and trustworthy policy enforcement between the edge IoT devices supported by the underlying blockchain properties e.g., scalability, auditability, and transparency. & Proposes a light-weight and decentralised secure access control framework for enforcing access control permissions using blockchain smart contracts. & Yes \\ [0.5ex]

\cite{DORRI2019180} & 2019 & Building a low-resourced consuming distributed time-based consensus algorithm to reduce processing overheads for verifying blocks of a blockchain network to provide fine-grained access control to large-scale IoT systems. & Proposes a `Light-weight Scalable Blockchain' solution for IoT systems called as LSB considering requirements of the low-resourced IoT devices and end-to-end security. & Yes \\ [0.5ex]

\cite{8964343} & 2020 & Examining the use of blockchain for access control and policy management for large-scale IoT systems. & Develops an access control framework called \textit{`Fabric-IoT'} that provides a decentralised, fine-grained and dynamic access control management for large-scale IoT systems supported by blockchain. & Yes \\ [0.5ex] 


\cite{9355161} & 2021 & Addressing the access control issues in IoT systems using blockchain over different functional domains.  & Proposes an attribute-managed access control framework for IoT using blockchain that supports cross-domain access requests. &  No \\ [0.5ex]

\cite{zhang-electronics9020285} & 2020 & Examining privacy-aware access control and efficient distribution and management of attributes using blockchain for large-scale IoT systems. & Proposes an attribute-based collaborative access control scheme using blockchain for IoT devices. &  Yes \\ [0.5ex]

\cite{almadhoun2018user} & 2018 & Examining how to enhance the scalability of an IoT system where the heavy computational takes are carried out by the fog nodes. & Proposes a framework that leverages flexibility to access control by considering resource constrained nature of the IoT devices supported by fog nodes. & Yes \\ [0.5ex]

\cite{DBLP-journals-048309} & 2019 & Examining data security in IoT access control using fog computing. & Proposes a blockchain-enabled fog architecture that provides a secure access control in IoT applications. & No \\ [0.5ex]

\cite{8308029} & 2017 &  Examining the use of fog nodes for heavy computational processing for IoT. & Develops a distributed and fine-grained access control architecture for IoT supported by fog computing. & Yes \\ [0.5ex] 

\hline
\end{tabular}
\label{tab:table-1}   
\end{table*}

\section{Discussion}
\label{open-questions}
Several proposals discuss the need for IoT access control and specific requirements and dedicated frameworks. That said, access control in IoT is application-specific and it requires dedicated infrastructures and contexts within which it will function. Table~\ref{tab:table-1} summarises the access control proposals for IoT based on blockchain that are discussed above based on their purposes, key contributions, and whether they are implemented.However, we have observed that there is no complete solution that can cover every aspect of an IoT system to determine secure access control by specifying its requirements~\cite{pal2020security}. Conventional access control mechanisms are not directly applicable in IoT due to its unique features, including resource constrained devices and dynamic characteristics. Access control requirements are different in various levels of IoT system at a fine-grained level. 

Blockchain has the potential to drive the access control issues in  IoT to address the limitations of the conventional solutions and to a higher extent to provide more flexibility, scalability, control, trust, and security \cite{WANG201910}. Due to specific requirements of IoT, the use of blockchain-based solutions needs to be focused on a particular issue and  infrastructure on which the access control is applied. Blockchain is immutable and auditable which facilitates access control management. The immutability of the blockchain makes it impossible for malicious nodes to modify access control rules. The auditability of the blockchain ensures that all interactions between nodes, and thus access to the data, is reflected  in blockchain. Thus, an attempt to access data or unauthorised access to the data will be recorded in the ledger which in turn enables the data/resource owner to detect the malicious activity and protect their resource~\cite{9358215}~\cite{khan2018iot}.


Among others, we have noted that the intrinsic features of blockchain helps in resource management by optimizing the resource utilisation rate and at the same time reducing the cost of the heavy-weight centralised service providers, transfer of access right from one entity to another entity, efficient enforcement of access control policies, management of attributes, and successively enhanced the scalability of fog nodes. These features are significant as access control decisions are highly dependent upon them. That said, blockchain can allow a platform for data collection and pass the collected data among IoT devices in a transparent and trustworthy fashion (by providing a verifiable and secure recording). For instance, with the provision of distributed computing and storage resource management for IoT access control using blockchain allows resource constraint IoT devices to perform local authorisation decisions in a faster way. In this, the authorisation chain can be verified inside the blockchain network in a decentralised manner and the access control conditions can be verified locally by the edge IoT devices~\cite{eberhardt2017or}.

\begin{table*}
\small
\caption {Comparison of access control requirements in blockchain-based access control and the commonly used access control mechanisms for the IoT systems. (AC = access control, BC = blockchain)}
\label{tab:comparison-requirements}    
\centering
\begin{tabular}{p{4.2cm} | p{1.7cm}p{1.7cm}p{1.7cm}p{1.7cm}}
\hline

    {AC Requirements} &{RBAC}&{ABAC}&{CapBAC}&{BC-based}\\ [0.5ex] \hline\hline

Scalability & Low & Medium & High & High \\ [0.5ex]
Ease of use & Medium & High & High & High \\ [0.5ex]
Choice of architecture & Centralised & Centralised & Distributed & Distributed \\ [0.5ex]
Data trust & High & Medium & Low & High \\ [0.5ex]
Continual control & Medium & High & High  & High \\ [0.5ex]
Security & High & Medium & Low & High \\ [0.5ex]
Support for integration & Yes & Yes & Yes & Yes \\ [0.5ex]
Cross domain access control & No & No & Yes & Yes \\ [0.5ex]

\hline 
\end{tabular}
\end{table*}

As noted above, using blockchain, this way of decentralised of access control can be an alternative for most of the centralised access control mechanisms, e.g., RBAC and ABAC. Recall, in RBAC and ABAC policy management and their informants are highly centralised and therefore it imposes several limitations when designing access control for a highly dynamic and scalable system like  IoT, including single point of failure and many to one traffic nature. In Table~\ref{tab:comparison-requirements}, we illustrate how  blockchain-based access control for IoT addresses  different salient features of access control requirements compared to the traditional RBAC, ABAC, and CapBAC mechanisms.

In Table~\ref{tab:comparison}, we illustrate the major advantages and disadvantages between blockchain and traditional access control mechanisms (e.g., RBAC, ABAC and CapBAC). However, based on the previous discussion, there are several open issues that must be addressed for a secure, efficient and robust access control for the IoT using blockchain-based systems in the future. We list them as follows:

\subsection{Policy Management}
In a highly dynamic and mobile system like the IoT, where there may be a large number of devices, being able to easily identify them, both uniquely and as groups and to provide access to the resources are challenging. Access control for the IoT requires the inclusion of proper policy enforcement mechanisms, which must define how the system should interact with other systems and entities~\cite{8968396}. We argue that this can be achieved by the underlying security policies determined by the policy decision component of the system. However, the challenge is how to place this policy decision component in a decentralised way to leverage the edge intelligence of the IoT devices~\cite{8109357}.


This enhances the specific requirements of the policy settings, including how policies are written and the corresponding permissions are granted~\cite{RAVIDAS201979}. For instance, in an IoT-enabled smart healthcare setting that is spread over multiple domains, cooperation and communication between different actors, clinical establishments and hospital management are crucial to share information seamlessly when communicating with one another. A significant research issue in this area is, how to manage access control policies in a flexible and fine-grained way for the IoT at scale without the need for a centralised component. This supports the use of blockchain for managing access control and efficient policy management in a decentralised way. There is a need for a policy management framework for blockchain to publish policies that can efficiently express the rights to access a resource and at the same time secure the distributed transfer of such access rights between the entities. In addition, managing access control policies in blockchain transforms the policy evaluation process to executable smart contracts which leverages several advantages of security features of reliability, traceability, and information integrity~\cite{8011213}.

\subsection{Trust Management}
The scale of the IoT systems makes it necessary to question certain aspects of traditional approaches to trust management~\cite{SHARMA2020475} \cite{voas2018internet}. However, managing trust in an IoT context is difficult. Heterogeneity and high dynamics in the systems bring new challenges to establish a trustworthy environment that further reinforces the absence of proper trust mechanisms in the IoT. It is important to quantify the perception of trust and the corresponding access control permissions that are context-sensitive, subjective, and in many cases vary in different ways based on the social issues~\cite{wu2021blockchain}. Contextual information must be also taken into consideration at the time of trust value verification and update, further complicating the situation~\cite{DOUGLAS2020100004}.

In IoT, the interaction between the entities may be for a very short period of time and the entities may be interacting only once for their entire life-time. In other cases, the interactions may be very frequent and may be for a longer period of time. Further, we note that the IoT devices are in general resource constrained. They have limited battery power, memory capacity and processing speed~\cite{101007-39393922}. Therefore, it is often not possible to store extensive interaction history or employ traditional heavy-weight security mechanisms in those devices for trust evaluation. It could be argued that, given the potentially highly unpredictable nature of interactions and uncertainty present in data (physical and digital data) that are coming from multiple sources in an IoT system, the need for trust in enforcing access control mechanisms is even greater in such systems~\cite{6513398} \cite{8328674} \cite{8250381}. That said, in many cases, employing a centralised trusted authority is not feasible in the IoT~\cite{Sabater2005}. 

In this context, a significant challenge is to develop a trust management framework for  IoT access control to deal with uncertainty present in such a decentralised system, e.g., multi-agent systems, where autonomous entities (e.g., agents) need to perform individual tasks, either for themselves or on behalf of a human user. In other words, given the desirable properties of trust (e.g., dynamic, content-dependent, transitive, etc.) and various sources of uncertainties, a major research question that is unsolved at present is how to represent trust for natural and artificial uncertainties. Natural uncertainty is about the outcome of a transaction, whereas artificial uncertainty results from second-hand experiences (i.e., opinions that are learned from other entities/sources). 
As noted above, among others, the potential adaptations of blockchain for  IoT can provide decentralised, secure and robust access control while replacing the centralised trusted authority~\cite{3205977-3205993}~\cite{3322431-3326327}~\cite{putra-9422738}. This will enable more transparency in IoT access control for both users and devices based on a level of collaboration, efficiency and improved visibility of trust among the entities.

\begin{table*}
\small
\caption {Major advantages and disadvantages of different access control mechanisms used for the IoT systems}
\label{tab:comparison}    
\centering
\begin{tabular}{p{2.5cm}p{6.1cm}p{6.1cm}}
\hline

{AC Mechanisms} & {{Advantages}} & {{ Disadvantages}} \\ [0.5ex] \hline \hline

RBAC & Provides stronger security by enforcing effective policy management. & Policy management and their informants are highly centralised. It typically requires explicit user assignment to specific roles and supports only pre-defined and static policies which do not support the scale of an IoT system. \\ [0.5ex] 

ABAC & Provides flexibility as the access control decision is performed based on attributes. It helps to enforce fine-grained access control policies in real-time. & In an IoT context, the use of ABAC raises important questions of the number of policy requirements, the policy evaluations, storage of attribute policy base, and the associated cost of applications. \\ [0.5ex]

CapBAC & Considers the resource-constrained characteristics of the IoT devices simplifies the distribution of permissions, and allows fine-grained access control. It is decentralised by nature. & Management of the number of capabilities that are required in a realistic IoT system and the issues of capability propagation and revocation are two common challenges when employing it for IoT systems.\\ [0.5ex]

Blockchain & Provides high security to prevent unauthorised data access. It is a distributed database of verifiable records. & Scalability remains an open issue for blockchain management. For instance, blockchain ledgers can expand large over time that raises the issue to download and store the ledger. \\ [0.5ex]
\hline

\end{tabular}
\end{table*}

\subsection{Standardisation}
The characteristics of IoT make it challenging to address interoperability and standardisation within the systems, applications, and various services~\cite{7821686}. In the IoT, there are many issues related to a diverse number of protocols  as  well as  the  lack  of agreement on  which  ones  work  best  for  individual layers  of an IoT architecture. Significantly, most of the Internet standards are designed for general purpose devices and do not account for the specific requirement of  IoT .  We observe that many of the access control models can fit in one integrated system but are unsuitable for others. Given the nature (e.g., resource-constrained, dynamic, highly scalable) of an IoT system, these access control requirements are varied, and enforcing a proper standardisation is extremely difficult for the IoT. This includes the potential disruption of the international communication lines, dynamically utilizing locally available IoT resources, and constant adaptation of new services and applications~\cite{7785888}. As such, a furture research issue is how to coordinate with different blockchain technologies, applications, and services to maintain seamless interoperability for the IoT systems in a standardised way. 

While various organisations, e.g., the IEEE and the Internet Engineering Task Force (IETF), are actively working on IoT standardisation, as of now, there is no standard platform for IoT aggregation whether it comes to applications and services~\cite{Pal18}. Today, many IoT devices are deployed with proprietary protocols which further makes it difficult to communicate among multiple IoT devices using a single blockchain platform~\cite{101007978555}.

\subsection{Identity Management}
We argue that the scope and nature of an IoT system mean that insisting on a definitive, unique, identity in every case is overly restrictive.  While in some circumstances such unique identification will be required, in other cases less defined identities will suffice for the needs of application functionality and policy specification. It is challenging to analyze, appraise and classify the various representations of identities in a detailed and comprehensive manner, and examine their suitability within the context of an IoT system~\cite{mahalle2010identity}. The vision of IoT implies that knowing the identities of individual entities before an interaction is, in many cases, impractical. To this end, future research should address how to design a flexible identity management framework for IoT access control at scale supported by blockchain technology~\cite{8726730}. 

There are several proposed identity management frameworks; however, they have not been demonstrated to adequately address the particular nature of the IoT, including its scale and heterogeneous context~\cite{pal2018modeling}. Blockchain can enable a robust network meshes in which IoT devices can communicate securely without the need for revealing each other’s identities. This can be achieved by employing efficient consensus mechanisms. We note that ABAC can be an alternative for managing IoT identity, where users are identified by sets of attributes e.g., name, age, location, etc~\cite{1011452903150-2911710} \cite{1011451866855-1866866}. Significantly, in ABAC, access control permissions are assigned based on the policies that are governed by the attributes. These attributes can be seen as the properties that describe specific features of users, resources, contexts, and conditions. The attributes of the user and those of the resource together determine the set of operations (based on the policies) that can be performed in a specific context. 
The use of attributes can provide a powerful method of specifying access policies in a flexible and fine-grained way that is particularly useful in an IoT system. Proposal~\cite{PAL201957} further demonstrates how using attributes, we can flexibly define IoT identity, which in turn specifies the available permissions. This reduces the number of policies that must be created by allowing a single attribute expression to provide access to multiple resources. Note that the same attributes may have different values in different contexts. For instance, let us consider the attribute `qualified' which may exist in two contexts e.g. `a taxi driver' and `a plane pilot'. In other words, in this case, the attribute `qualified' may have the value `true' in the context of a taxi driver but `not true' in the case of a plane pilot. 
There are also possibilities to create smart contracts to enable controlled data disclosure which in turn help in IoT identity management providing `trusted' auditability required for IoT access control systems~\cite{zhu2018identity}~\cite{8726747}~\cite{bouras2021lightweight}. 


\subsection{Road Ahead}
We observed that there is a need for protecting IoT systems from unauthorised users, services, and applications by enforcing appropriate access control mechanisms that satisfy the various characteristics and requirements of an IoT system. This could not be fixed with the use of simple software patches or applying heavy-weight security mechanisms inside the resource constrained IoT devices, rather it requires the dedicated access control architecture, light-weight security mechanisms, secure communication protocols and appropriate policy management~\cite{9323061}. Moreover, the scope and variety of recent technological developments impose sophisticated constraints e.g., find-grained security attributes, for authentication and authorisation in IoT systems that are not supported by earlier security frameworks. Entities in IoT still need some basis on which to determine whether to interact with one another, including the bestowal and acknowledgement of access control rights. Consider the context of the size and scale in an IoT system, the access control mechanisms must be asynchronous. Using blockchain access control policies are encrypted and securely stored inside the blocks that gives more flexibility in accessing these policies when required. That said, it should not be assumed that entities in an IoT system are constantly in communication with each or the wider system~\cite{9013321}. The use of blockchain also reduces the dependencies over a single central manager. A single control point would run the risks of single point of failure, amongst others.
We also noted that there is a need to minimize the overhead created by many security mechanisms for the IoT. These mechanisms can be enforced in various ways in access control and policy management e.g., in back-end management, secure design, and development practices, or even at an application level. Recall, access control in IoT is application-specific and it requires dedicated infrastructures and contexts within which it will function~\cite{8592253}.

We noted, among others, that building trust between various entities in a highly dynamic and scalable system like the IoT is a challenging task that is significant when sharing information and controlling their access. Most of the access control approaches at present enhance the traditional distributed trust management systems for the IoT, lacking the proper need for trust in an IoT context~\cite{9422738}. 
We argue that there is a need to investigate the fusion of multiple observations along with the distributed aspects of trust in decision making under multiple-sources for large-scale IoT systems. We note that comprehensive research in IoT trust is required which must consider a complete view of an IoT system, along with its access control needs and design requirements supported by blockchain technology. 

Moreover, the convergence of Artificial Intelligence (AI) in blockchain technology is promising and can be combined in multiple dimensions to build a sustainable access control mechanism for IoT systems~\cite{9382024}. With the help of Machine Learning (ML) algorithms, AI can improve access control processes by detecting access patterns and optimize the network’s maintenance with little or no human interference. However, a broad spectrum of adaptation and integration of blockchain and AI technologies in IoT bring both opportunities and challenges in terms of scalability, security, identity, interoperability, and trust management. For blockchain transformation in AI need to provide accuracy and certainty to devise consensus protocols for mining nodes in developing successful cognitive systems. Some of the challenges for ML adaptation in blockchain in IoT include the design and development of smart agents with improved learning capability to regulate blockchain, and the proper consideration of uncertainty, e.g., the risks related to a given access control mechanism,  that are present in such highly dynamic IoT systems. An efficient access control framework must be built to combine ML-assisted data fusion mechanisms for multi-layer and multi-vendor blockchain systems for data authorisation in IoT. The other aspect could be the implementation of ML algorithms in smart contracts~\cite{hussain2020machine}. This requires novel solutions from both the theoretical and practical sides. There is a need for methodological, algorithmic, mathematical, and computational models combining blockchain and AI to solve the theoretical and practical problems in large-scale IoT systems~\cite{salah2019blockchain}.

\section{Conclusion}
\label{conclusion}
With the rapid development of the IoT, research in the IoT landscape becomes an important issue and it continues to grow every day. However, one significant challenge is the provision of security within the IoT, in particular, the need for access control is paramount. We have noted that the traditional access control mechanisms cannot achieve efficient management of access control policies and enforcement of authorisation decisions for large-scale IoT systems. That said, IoT systems require a unique solution for access control. We observed that the blockchain has the potential to achieve the several desire goals of IoT access control e.g., tamper-proof, trusted, decentralised control, data transparency, and auditability. However, one common issue to the existing access control surveys in IoT is that they have a lack of focus on the emerging blockchain technology. In this paper, we reviewed recent trends of such blockchain-based access control solutions for the IoT systems. We provided a systematic discussion of the existing blockchain-based access control solutions for the IoT by categorizing them based on the certain access control needs they satisfy e.g., resource management, access rights transfer, permission enforcement, attribute management, and scalability. We have shown how blockchain can improve the limitations of IoT access control issues over the traditional access control mechanisms addressing these access control needs. Our review also explored a list of important future research directions in order to address policy management, trust, identity management, and inclusion of data-driven technology (e.g., artificial intelligence) to deliver a flexible, decentralised, trustworthy and fine-grained access control for the IoT supported by blockchain.

\ifCLASSOPTIONcaptionsoff
  \newpage
\fi

\bibliographystyle{IEEEtran}
\bibliography{mybibfile-12}

\begin{thebibliography}{100}
\providecommand{\url}[1]{#1}
\csname url@samestyle\endcsname
\providecommand{\newblock}{\relax}
\providecommand{\bibinfo}[2]{#2}
\providecommand{\BIBentrySTDinterwordspacing}{\spaceskip=0pt\relax}
\providecommand{\BIBentryALTinterwordstretchfactor}{4}
\providecommand{\BIBentryALTinterwordspacing}{\spaceskip=\fontdimen2\font plus
\BIBentryALTinterwordstretchfactor\fontdimen3\font minus
  \fontdimen4\font\relax}
\providecommand{\BIBforeignlanguage}[2]{{%
\expandafter\ifx\csname l@#1\endcsname\relax
\typeout{** WARNING: IEEEtran.bst: No hyphenation pattern has been}%
\typeout{** loaded for the language `#1'. Using the pattern for}%
\typeout{** the default language instead.}%
\else
\language=\csname l@#1\endcsname
\fi
#2}}
\providecommand{\BIBdecl}{\relax}
\BIBdecl

\bibitem{8286847}
T.~Qiu, N.~Chen, K.~Li, M.~Atiquzzaman, and W.~Zhao, ``How can heterogeneous
  internet of things build our future: A survey,'' \emph{IEEE Communications
  Surveys Tutorials}, vol.~20, no.~3, pp. 2011--2027, thirdquarter 2018.

\bibitem{cisco-2021}
``Cisco: The zettabyte era: Trends and analysis,''
  \url{https://www.cisco.com/c/en/us/solutions/collateral/service-provider/visual-networking-index-vni/vni-hyperconnectivity-wp.html},
  [Online: accessed 24-Oct-2018].

\bibitem{lin:iot-security-privacy-survey-2017}
\BIBentryALTinterwordspacing
J.~Lin, W.~Yu, N.~Zhang, X.~Yang, H.~Zhang, and W.~Zhao, ``{A Survey on
  Internet of Things: Architecture, Enabling Technologies, Security and
  Privacy, and Applications},'' \emph{IEEE Internet of Things Journal}, vol.~4,
  no.~5, pp. 1125--1142, Oct. 2017. [Online]. Available:
  \url{http://dx.doi.org/10.1109/jiot.2017.2683200}
\BIBentrySTDinterwordspacing

\bibitem{alqassem:iot-security-requirements-2014}
\BIBentryALTinterwordspacing
I.~Alqassem, ``{Privacy and security requirements framework for the internet of
  things (IoT)}.'' [Online]. Available:
  \url{https://dl.acm.org/citation.cfm?id=2591201}
\BIBentrySTDinterwordspacing

\bibitem{8086136}
M.~Frustaci, P.~Pace, G.~Aloi, and G.~Fortino, ``Evaluating critical security
  issues of the iot world: Present and future challenges,'' \emph{IEEE Internet
  of Things Journal}, vol.~5, no.~4, pp. 2483--2495, Aug 2018.

\bibitem{ANDALOUSSI20181031}
\BIBentryALTinterwordspacing
Y.~Andaloussi, M.~E. Ouadghiri], Y.~Maurel, J.~Bonnin, and H.~Chaoui, ``Access
  control in iot environments: Feasible scenarios,'' \emph{Procedia Computer
  Science}, vol. 130, pp. 1031--1036, 2018, the 9th International Conference
  on Ambient Systems, Networks and Technologies (ANT 2018). [Online]. Available:
  \url{http://www.sciencedirect.com/science/article/pii/S1877050918305064}
\BIBentrySTDinterwordspacing

\bibitem{tolone2005access}
W.~Tolone, G.-J. Ahn, T.~Pai, and S.-P. Hong, ``Access control in collaborative
  systems,'' \emph{ACM Computing Surveys (CSUR)}, vol.~37, no.~1, pp. 29--41,
  2005.

\bibitem{dabbagh:iot-security-2017}
\BIBentryALTinterwordspacing
M.~Dabbagh and A.~Rayes, ``{Internet of Things Security and Privacy},'' in
  \emph{Internet of Things From Hype to Reality}.\hskip 1em plus 0.5em minus
  0.4em\relax Springer International Publishing, 2017, pp. 195--223. [Online].
  Available: \url{http://dx.doi.org/10.1007/978-3-319-44860-2\_8}
\BIBentrySTDinterwordspacing

\bibitem{pal2019policy}
S.~Pal, M.~Hitchens, V.~Varadharajan, and T.~Rabehaja, ``Policy-based access
  control for constrained healthcare resources in the context of the internet
  of things,'' \emph{Journal of Network and Computer Applications}, vol. 139,
  pp. 57--74, 2019.

\bibitem{MAJEED2021103007}
\BIBentryALTinterwordspacing
U.~Majeed, L.~U. Khan, I.~Yaqoob, S.~A. Kazmi, K.~Salah, and C.~S. Hong,
  ``Blockchain for iot-based smart cities: Recent advances, requirements, and
  future challenges,'' \emph{Journal of Network and Computer Applications},
  vol. 181, p. 103007, 2021. [Online]. Available:
  \url{https://www.sciencedirect.com/science/article/pii/S1084804521000345}
\BIBentrySTDinterwordspacing

\bibitem{wang2019survey}
X.~Wang, X.~Zha, W.~Ni, R.~P. Liu, Y.~J. Guo, X.~Niu, and K.~Zheng, ``Survey on
  blockchain for internet of things,'' \emph{Computer Communications}, vol.
  136, pp. 10--29, 2019.

\bibitem{101145-3180457-3180458}
\BIBentryALTinterwordspacing
C.~Dukkipati, Y.~Zhang, and L.~C. Cheng, ``Decentralized, blockchain based
  access control framework for the heterogeneous internet of things,'' in
  \emph{Proceedings of the Third ACM Workshop on Attribute-Based Access
  Control}, ser. ABAC'18.\hskip 1em plus 0.5em minus 0.4em\relax New York, NY,
  USA: Association for Computing Machinery, 2018, p. 61–69. [Online].
  Available: \url{https://doi.org/10.1145/3180457.3180458}
\BIBentrySTDinterwordspacing

\bibitem{101145-3350546-3352561}
\BIBentryALTinterwordspacing
S.~Rouhani and R.~Deters, ``Blockchain based access control systems: State of
  the art and challenges,'' in \emph{IEEE/WIC/ACM International Conference on
  Web Intelligence}, ser. WI'19.\hskip 1em plus 0.5em minus 0.4em\relax New
  York, NY, USA: Association for Computing Machinery, 2019, p. 423–428.
  [Online]. Available: \url{https://doi.org/10.1145/3350546.3352561}
\BIBentrySTDinterwordspacing

\bibitem{fotiou2016access}
N.~Fotiou, T.~Kotsonis, G.~F. Marias, and G.~C. Polyzos, ``Access control for
  the internet of things,'' in \emph{Secure Internet of Things (SIoT), 2016
  International Workshop on}.\hskip 1em plus 0.5em minus 0.4em\relax IEEE,
  2016, pp. 29--38.

\bibitem{elsayed2016access}
W.~Elsayed, T.~Gaber, N.~Zhang, and M.~I. Moussa, ``Access control models for
  pervasive environments: A survey,'' in \emph{The 1st International Conference
  on Advanced Intelligent System and Informatics (AISI2015), November 28-30,
  2015, Beni Suef, Egypt}.\hskip 1em plus 0.5em minus 0.4em\relax Springer,
  2016, pp. 511--522.

\bibitem{ranjan:iot-access-control-survey-2016}
\BIBentryALTinterwordspacing
A.~Ranjan and G.~Somani, ``{Access Control and Authentication in the Internet
  of Things Environment},'' in \emph{Connectivity Frameworks for Smart
  Devices}, ser. Computer Communications and Networks, Z.~Mahmood, Ed.\hskip
  1em plus 0.5em minus 0.4em\relax Springer International Publishing, 2016, pp.
  283--305. [Online]. Available:
  \url{http://dx.doi.org/10.1007/978-3-319-33124-9\_12}
\BIBentrySTDinterwordspacing

\bibitem{zhang2016access}
Y.~Zhang and X.~Wu, ``Access control in internet of things: A survey,''
  \emph{arXiv preprint arXiv:1610.01065}, 2016.

\bibitem{8038503}
M.~Alramadhan and K.~Sha, ``An overview of access control mechanisms for
  internet of things,'' in \emph{2017 26th International Conference on Computer
  Communication and Networks (ICCCN)}, July 2017, pp. 1--6.

\bibitem{RAVIDAS201979}
\BIBentryALTinterwordspacing
S.~Ravidas, A.~Lekidis, F.~Paci, and N.~Zannone, ``Access control in
  internet-of-things: A survey,'' \emph{Journal of Network and Computer
  Applications}, vol. 144, pp. 79 -- 101, 2019. [Online]. Available:
  \url{http://www.sciencedirect.com/science/article/pii/S108480451930222X}
\BIBentrySTDinterwordspacing

\bibitem{Bertin2019}
\BIBentryALTinterwordspacing
E.~Bertin, D.~Hussein, C.~Sengul, and V.~Frey, ``Access control in the internet
  of things: a survey of existing approaches and open research questions,''
  \emph{Annals of Telecommunications}, Mar 2019. [Online]. Available:
  \url{https://doi.org/10.1007/s12243-019-00709-7}
\BIBentrySTDinterwordspacing

\bibitem{ouaddah:iot-access-challenges-2017}
\BIBentryALTinterwordspacing
A.~Ouaddah, H.~Mousannif, A.~Abou~Elkalam, and A.~Ait~Ouahman, ``{Access
  control in the Internet of Things: Big challenges and new opportunities},''
  \emph{Computer Networks}, vol. 112, pp. 237--262, Jan. 2017. [Online].
  Available: \url{http://dx.doi.org/10.1016/j.comnet.2016.11.007}
\BIBentrySTDinterwordspacing

\bibitem{riabi-8766453}
I.~Riabi, H.~K.~B. Ayed, and L.~A. Saidane, ``A survey on blockchain based
  access control for internet of things,'' in \emph{2019 15th International
  Wireless Communications Mobile Computing Conference (IWCMC)}, 2019, pp.
  502--507.

\bibitem{343113734566789}
C.~Amritanand and P.~Vipin, ``A survey on blockchain based access control for
  iot,'' in \emph{Proceedings of International Conference on Recent Trends in
  Computing, Communication and Networking Technologies (ICRTCCNT)}, 2019, pp.
  618--623.
\BIBentrySTDinterwordspacing

\bibitem{corr-abs-1908-08503}
\BIBentryALTinterwordspacing
S.~Rouhani and R.~Deters, ``Blockchain based access control systems: State of
  the art and challenges,'' \emph{CoRR}, vol. abs/1908.08503, 2019. [Online].
  Available: \url{http://arxiv.org/abs/1908.08503}
\BIBentrySTDinterwordspacing

\bibitem{sym12101663}
\BIBentryALTinterwordspacing
A.~I. Abdi, F.~E. Eassa, K.~Jambi, K.~Almarhabi, and A.~S. A.-M. AL-Ghamdi,
  ``Blockchain platforms and access control classification for iot systems,''
  \emph{Symmetry}, vol.~12, no.~10, 2020. [Online]. Available:
  \url{https://www.mdpi.com/2073-8994/12/10/1663}
\BIBentrySTDinterwordspacing

\bibitem{9223297}
F.~Ghaffari, E.~Bertin, J.~Hatin, and N.~Crespi, ``Authentication and access
  control based on distributed ledger technology: A survey,'' in \emph{2020 2nd
  Conference on Blockchain Research Applications for Innovative Networks and
  Services (BRAINS)}, 2020, pp. 79--86.

\bibitem{8968396}
J.~{Qiu}, Z.~{Tian}, C.~{Du}, Q.~{Zuo}, S.~{Su}, and B.~{Fang}, ``A survey on
  access control in the age of internet of things,'' \emph{IEEE Internet of
  Things Journal}, vol.~7, no.~6, pp. 4682--4696, 2020.

\bibitem{8340813}
F.~Jameel, Z.~Hamid, F.~Jabeen, S.~Zeadally, and M.~A. Javed, ``A survey of
  device-to-device communications: Research issues and challenges,'' \emph{IEEE
  Communications Surveys Tutorials}, vol.~20, no.~3, pp. 2133--2168,
  thirdquarter 2018.

\bibitem{kevin-ashton:iot-2009}
\BIBentryALTinterwordspacing
K.~Ashton, ``{That 'Internet of Things' Thing},'' \emph{RFID}, 2009. [Online].
  Available: \url{http://www.rfidjournal.com/articles/view?4986}
\BIBentrySTDinterwordspacing

\bibitem{8489954}
I.~Makhdoom, M.~Abolhasan, J.~Lipman, R.~P. Liu, and W.~Ni, ``Anatomy of
  threats to the internet of things,'' \emph{IEEE Communications Surveys
  Tutorials}, pp. 1--1, 2018.

\bibitem{8470752}
V.~{Beltran} and A.~F. {Skarmeta}, ``Overview of device access control in the
  iot and its challenges,'' \emph{IEEE Communications Magazine}, vol.~57,
  no.~1, pp. 154--160, January 2019.

\bibitem{8993839}
Y.~A. {Qadri}, A.~{Nauman}, Y.~B. {Zikria}, A.~V. {Vasilakos}, and S.~W. {Kim},
  ``The future of healthcare internet of things: A survey of emerging
  technologies,'' \emph{IEEE Communications Surveys Tutorials}, vol.~22, no.~2,
  pp. 1121--1167, 2020.

\bibitem{wang:iot-health}
\BIBentryALTinterwordspacing
W.~Wang, J.~Li, L.~Wang, and W.~Zhao, ``{The internet of things for resident
  health information service platform research},'' in \emph{IET International
  Conference on Communication Technology and Application (ICCTA)}.\hskip 1em
  plus 0.5em minus 0.4em\relax IET, Oct. 2011, pp. 631--635. [Online].
  Available: \url{http://dx.doi.org/10.1049/cp.2011.0745}
\BIBentrySTDinterwordspacing

\bibitem{101145-31444573144485}
\BIBentryALTinterwordspacing
S.~Pal, M.~Hitchens, V.~Varadharajan, and T.~Rabehaja, ``On design of a
  fine-grained access control architecture for securing iot-enabled smart
  healthcare systems,'' in \emph{Proceedings of the 14th EAI International
  Conference on Mobile and Ubiquitous Systems: Computing, Networking and
  Services}, ser. MobiQuitous 2017.\hskip 1em plus 0.5em minus 0.4em\relax New
  York, NY, USA: Association for Computing Machinery, 2017, p. 432–441.
  [Online]. Available: \url{https://doi.org/10.1145/3144457.3144485}
\BIBentrySTDinterwordspacing

\bibitem{chahid:iot-security-2017}
\BIBentryALTinterwordspacing
Y.~Chahid, M.~Benabdellah, and A.~Azizi, ``{Internet of things security},'' in
  \emph{2017 International Conference on Wireless Technologies, Embedded and
  Intelligent Systems (WITS)}.\hskip 1em plus 0.5em minus 0.4em\relax IEEE,
  Apr. 2017, pp. 1--6. [Online]. Available:
  \url{http://dx.doi.org/10.1109/wits.2017.7934655}
\BIBentrySTDinterwordspacing

\bibitem{8716500}
M.~Alaslani, F.~Nawab, and B.~Shihada, ``Blockchain in iot systems: End-to-end
  delay evaluation,'' \emph{IEEE Internet of Things Journal}, vol.~6, no.~5,
  pp. 8332--8344, 2019.

\bibitem{8897627}
I.~{Butun}, P.~{Österberg}, and H.~{Song}, ``Security of the internet of
  things: Vulnerabilities, attacks, and countermeasures,'' \emph{IEEE
  Communications Surveys Tutorials}, vol.~22, no.~1, pp. 616--644, 2020.

\bibitem{Patel2019}
\BIBentryALTinterwordspacing
C.~Patel and N.~Doshi, \emph{Security Challenges in IoT Cyber World}.\hskip 1em
  plus 0.5em minus 0.4em\relax Cham: Springer International Publishing, 2019,
  pp. 171--191. [Online]. Available:
  \url{https://doi.org/10.1007/978-3-030-01560-2_8}
\BIBentrySTDinterwordspacing

\bibitem{mahler:know-your-enemy-iot-2018}
\BIBentryALTinterwordspacing
T.~Mahler, N.~Nissim, E.~Shalom, I.~Goldenberg, G.~Hassman, A.~Makori,
  T.~Kochav, U.~Elovici, and Y.~Shahar, ``{Know Your Enemy: Characteristics of
  Cyber-Attacks on Medical Imaging Devices},'' Feb. 2018. [Online]. Available:
  \url{http://arxiv.org/abs/1801.05583}
\BIBentrySTDinterwordspacing

\bibitem{sun2018security}
\BIBentryALTinterwordspacing
W.~Sun, Z.~Cai, Y.~Li, F.~Liu, S.~Fang, and G.~Wang, ``Security and privacy in
  the medical internet of things: A review,'' \emph{Security and Communication
  Networks}, vol. 2018, 2018. [Online]. Available:
  \url{https://doi.org/10.1155/2018/5978636}
\BIBentrySTDinterwordspacing

\bibitem{wired-iot-hack-2015}
WIRED, ``{How the Internet of Things got Hacked},''
  \url{https://www.wired.com/2015/12/2015-the-year-the-internet-of-things-got-hacked/},
  2015, [Online: accessed 01-Oct-2017].

\bibitem{mirai-2016}
``{Mirai Botnet DDoS Attack Type},''
  \url{https://www.corero.com/resources/ddos-attack-types/mirai-botnet-ddos-attack.html/},
  2016, [Online: accessed 10-Oct-2018].

\bibitem{doll-hack-2015}
``{The Internet of Things: Cayla doll is banned in Germany over privacy and
  security concerns},''
  \url{https://www.lexology.com/library/detail.aspx?g=d3a5448e-ecbc-41fb-b0cb-3d28bdfe841e},
  2017, [Online: accessed 01-Oct-2018].

\bibitem{findland-2016}
\BIBentryALTinterwordspacing
S.~Robinson, ``Smart home attacks are a reality, even as the smart home market
  soars,'' 2019, [Accessed 25-May-2019]. [Online]. Available:
  \url{https://www.cisco.com/c/en/us/solutions/internet-of-things/smart-home-attacks.html}
\BIBentrySTDinterwordspacing

\bibitem{uganya2021survey}
G.~Uganya, Radhika, and N.~Vijayaraj, ``A survey on internet of things:
  Applications, recent issues, attacks, and security mechanisms,''
  \emph{Journal of Circuits, Systems and Computers}, vol.~30, no.~05, p.
  2130006, 2021.

\bibitem{kouicem2018internet}
D.~E. Kouicem, A.~Bouabdallah, and H.~Lakhlef, ``Internet of things security: A
  top-down survey,'' \emph{Computer Networks}, 2018.

\bibitem{ezema2018open}
E.~Ezema, A.~Abdullah, and N.~F.~B. Mohd, ``Open issues and security challenges
  of data communication channels in distributed internet of things (iot): A
  survey,'' 2018.

\bibitem{conti2018internet}
M.~Conti, A.~Dehghantanha, K.~Franke, and S.~Watson, ``Internet of things
  security and forensics: Challenges and opportunities,'' 2018.

\bibitem{8058363}
J.~Deogirikar and A.~Vidhate, ``Security attacks in iot: A survey,'' in
  \emph{2017 International Conference on I-SMAC (IoT in Social, Mobile,
  Analytics and Cloud) (I-SMAC)}, Feb 2017, pp. 32--37.

\bibitem{8029379}
Z.~Zheng, S.~Xie, H.~Dai, X.~Chen, and H.~Wang, ``An overview of blockchain
  technology: Architecture, consensus, and future trends,'' in \emph{2017 IEEE
  International Congress on Big Data (BigData Congress)}, June 2017, pp.
  557--564.

\bibitem{8123011}
D.~Mingxiao, M.~Xiaofeng, Z.~Zhe, W.~Xiangwei, and C.~Qijun, ``A review on
  consensus algorithm of blockchain,'' in \emph{2017 IEEE International
  Conference on Systems, Man, and Cybernetics (SMC)}, 2017, pp. 2567--2572.

\bibitem{7945805}
M.~Conoscenti, A.~Vetrò, and J.~C.~D. Martin, ``Blockchain for the internet of
  things: A systematic literature review,'' in \emph{2016 IEEE/ACS 13th
  International Conference of Computer Systems and Applications (AICCSA)}, Nov
  2016, pp. 1--6.

\bibitem{bosch}
\BIBentryALTinterwordspacing
``Bosch connected devices and solutions,'' 2019. [Online]. Available:
  \url{https://xdk.bosch-connectivity.com/}
\BIBentrySTDinterwordspacing

\bibitem{hdac}
\BIBentryALTinterwordspacing
``Hyundai digital asset company,'' 2019. [Online]. Available:
  \url{https://www.hdactech.com/en/index.do}
\BIBentrySTDinterwordspacing

\bibitem{8370027}
T.~M. Fernández-Caramés and P.~Fraga-Lamas, ``A review on the use of
  blockchain for the internet of things,'' \emph{IEEE Access}, vol.~6, pp.
  32\,979--33\,001, 2018.

\bibitem{KHAN2018395}
\BIBentryALTinterwordspacing
M.~A. Khan and K.~Salah, ``Iot security: Review, blockchain solutions, and open
  challenges,'' \emph{Future Generation Computer Systems}, vol.~82, pp. 395 --
  411, 2018. [Online]. Available:
  \url{http://www.sciencedirect.com/science/article/pii/S0167739X17315765}
\BIBentrySTDinterwordspacing

\bibitem{DBLP-journals-07448}
\BIBentryALTinterwordspacing
F.~Restuccia, S.~D'Oro, S.~S. Kanhere, T.~Melodia, and S.~K. Das, ``Blockchain
  for the internet of things: Present and future,'' \emph{CoRR}, vol.
  abs/1903.07448, 2019. [Online]. Available:
  \url{http://arxiv.org/abs/1903.07448}
\BIBentrySTDinterwordspacing

\bibitem{s18082575}
\BIBentryALTinterwordspacing
A.~Panarello, N.~Tapas, G.~Merlino, F.~Longo, and A.~Puliafito, ``Blockchain
  and iot integration: A systematic survey,'' \emph{Sensors}, vol.~18, no.~8,
  2018. [Online]. Available: \url{http://www.mdpi.com/1424-8220/18/8/2575}
\BIBentrySTDinterwordspacing

\bibitem{REYNA2018173}
\BIBentryALTinterwordspacing
A.~Reyna, C.~Martín, J.~Chen, E.~Soler, and M.~Díaz, ``On blockchain and its
  integration with iot. challenges and opportunities,'' \emph{Future Generation
  Computer Systems}, vol.~88, pp. 173 -- 190, 2018. [Online]. Available:
  \url{http://www.sciencedirect.com/science/article/pii/S0167739X17329205}
\BIBentrySTDinterwordspacing

\bibitem{8580364}
M.~S. Ali, M.~Vecchio, M.~Pincheira, K.~Dolui, F.~Antonelli, and M.~H. Rehmani,
  ``Applications of blockchains in the internet of things: A comprehensive
  survey,'' \emph{IEEE Communications Surveys Tutorials}, pp. 1--1, 2018.

\bibitem{dedeoglu2020blockchain}
V.~Dedeoglu, R.~Jurdak, A.~Dorri, R.~Lunardi, R.~Michelin, A.~Zorzo, and
  S.~Kanhere, ``Blockchain technologies for iot,'' in \emph{Advanced
  Applications of Blockchain Technology}.\hskip 1em plus 0.5em minus
  0.4em\relax Springer, 2020, pp. 55--89.

\bibitem{7917634}
A.~Dorri, S.~S. Kanhere, R.~Jurdak, and P.~Gauravaram, ``Blockchain for iot
  security and privacy: The case study of a smart home,'' in \emph{2017 IEEE
  International Conference on Pervasive Computing and Communications Workshops
  (PerCom Workshops)}, 2017, pp. 618--623.

\bibitem{8306880}
O.~Novo, ``Blockchain meets iot: An architecture for scalable access management
  in iot,'' \emph{IEEE Internet of Things Journal}, vol.~5, no.~2, pp.
  1184--1195, April 2018.

\bibitem{8386853}
Y.~{Zhang}, S.~{Kasahara}, Y.~{Shen}, X.~{Jiang}, and J.~{Wan}, ``Smart
  contract-based access control for the internet of things,'' \emph{IEEE
  Internet of Things Journal}, pp. 1--1, 2019.

\bibitem{10-1007-978-3-319}
M.~Nuss, A.~Puchta, and M.~Kunz, ``Towards blockchain-based identity and access
  management for internet of things in enterprises,'' in \emph{Trust, Privacy
  and Security in Digital Business}, S.~Furnell, H.~Mouratidis, and G.~Pernul,
  Eds.\hskip 1em plus 0.5em minus 0.4em\relax Cham: Springer International
  Publishing, 2018, pp. 167--181.

\bibitem{DBLP-1804-09267}
\BIBentryALTinterwordspacing
R.~Xu, Y.~Chen, E.~Blasch, and G.~Chen, ``Blendcac: {A} blockchain-enabled
  decentralized capability-based access control for iots,'' \emph{CoRR}, vol.
  abs/1804.09267, 2018. [Online]. Available:
  \url{http://arxiv.org/abs/1804.09267}
\BIBentrySTDinterwordspacing

\bibitem{8421332}
T.~Le and M.~W. Mutka, ``Capchain: A privacy preserving access control
  framework based on blockchain for pervasive environments,'' in \emph{2018
  IEEE International Conference on Smart Computing (SMARTCOMP)}, June 2018, pp.
  57--64.

\bibitem{8752021}
S.~{Pal}, T.~{Rabehaja}, M.~{Hitchens}, V.~{Varadharajan}, and A.~{Hill}, ``On
  the design of a flexible delegation model for the internet of things using
  blockchain,'' \emph{IEEE Transactions on Industrial Informatics}, vol.~16,
  no.~5, pp. 3521--3530, 2020.

\bibitem{pal-iot-8894097}
S.~Pal, T.~Rabehaja, A.~Hill, M.~Hitchens, and V.~Varadharajan, ``On the
  integration of blockchain to the internet of things for enabling access right
  delegation,'' \emph{IEEE Internet of Things Journal}, vol.~7, no.~4, pp.
  2630--2639, 2020.

\bibitem{ALI2019}
\BIBentryALTinterwordspacing
G.~Ali, N.~Ahmad, Y.~Cao, M.~Asif, H.~Cruickshank, and Q.~E. Ali, ``Blockchain
  based permission delegation and access control in internet of things
  (baci),'' \emph{Computers \& Security}, 2019. [Online]. Available:
  \url{http://www.sciencedirect.com/science/article/pii/S0167404819301208}
\BIBentrySTDinterwordspacing

\bibitem{Shafagh-2017-TBA}
\BIBentryALTinterwordspacing
H.~Shafagh, L.~Burkhalter, A.~Hithnawi, and S.~Duquennoy, ``Towards
  blockchain-based auditable storage and sharing of iot data,'' in
  \emph{Proceedings of the 2017 on Cloud Computing Security Workshop}, ser.
  CCSW '17.\hskip 1em plus 0.5em minus 0.4em\relax New York, NY, USA: ACM,
  2017, pp. 45--50. [Online]. Available:
  \url{http://doi.acm.org/10.1145/3140649.3140656}
\BIBentrySTDinterwordspacing

\bibitem{algarni2021blockchain}
S.~Algarni, F.~Eassa, K.~Almarhabi, A.~Almalaise, E.~Albassam, K.~Alsubhi, and
  M.~Yamin, ``Blockchain-based secured access control in an iot system,''
  \emph{Applied Sciences}, vol.~11, no.~4, p. 1772, 2021.

\bibitem{DORRI2019180}
\BIBentryALTinterwordspacing
A.~Dorri, S.~S. Kanhere, R.~Jurdak, and P.~Gauravaram, ``Lsb: A lightweight
  scalable blockchain for iot security and anonymity,'' \emph{Journal of
  Parallel and Distributed Computing}, vol. 134, pp. 180--197, 2019. [Online].
  Available:
  \url{https://www.sciencedirect.com/science/article/pii/S0743731518307688}
\BIBentrySTDinterwordspacing

\bibitem{8964343}
H.~{Liu}, D.~{Han}, and D.~{Li}, ``Fabric-iot: A blockchain-based access
  control system in iot,'' \emph{IEEE Access}, vol.~8, pp. 18\,207--18\,218,
  2020.

\bibitem{9355161}
S.~{Sun}, R.~{Du}, S.~{Chen}, and W.~{Li}, ``Blockchain-based iot access
  control system: Towards security, lightweight, and cross-domain,'' \emph{IEEE
  Access}, vol.~9, pp. 36\,868--36\,878, 2021.

\bibitem{zhang-electronics9020285}
\BIBentryALTinterwordspacing
Y.~Zhang, B.~Li, B.~Liu, J.~Wu, Y.~Wang, and X.~Yang, ``An attribute-based
  collaborative access control scheme using blockchain for iot devices,''
  \emph{Electronics}, vol.~9, no.~2, 2020. [Online]. Available:
  \url{https://www.mdpi.com/2079-9292/9/2/285}
\BIBentrySTDinterwordspacing

\bibitem{almadhoun2018user}
R.~Almadhoun, M.~Kadadha, M.~Alhemeiri, M.~Alshehhi, and K.~Salah, ``A user
  authentication scheme of iot devices using blockchain-enabled fog nodes,'' in
  \emph{2018 IEEE-ACS 15th International Conference on Computer Systems and
  Applications (AICCSA)}.\hskip 1em plus 0.5em minus 0.4em\relax IEEE, 2018,
  pp. 1--8.

\bibitem{DBLP-journals-048309}
\BIBentryALTinterwordspacing
M.~Farhadi, D.~Miorandi, and G.~Pierre, ``Blockchain enabled fog structure to
  provide data security in iot applications,'' \emph{CoRR}, vol.
  abs/1901.04830, 2019. [Online]. Available:
  \url{http://arxiv.org/abs/1901.04830}
\BIBentrySTDinterwordspacing

\bibitem{8308029}
I.~{Riabi}, L.~A. {Saidane}, and H.~K. {Ayed}, ``A proposal of a distributed
  access control over fog computing: The its use case,'' in \emph{2017
  International Conference on Performance Evaluation and Modeling in Wired and
  Wireless Networks (PEMWN)}, Nov 2017, pp. 1--7.

\bibitem{hameed2019understanding}
S.~Hameed, F.~I. Khan, and B.~Hameed, ``Understanding security requirements and
  challenges in internet of things (iot): A review,'' \emph{Journal of Computer
  Networks and Communications}, vol. 2019, 2019.

\bibitem{unal2020policy}
D.~Unal, M.~Hammoudeh, and M.~S. Kiraz, ``Policy specification and verification
  for blockchain and smart contracts in 5g networks,'' \emph{ICT Express},
  vol.~6, no.~1, pp. 43--47, 2020.

\bibitem{Kiwelekar2021}
\BIBentryALTinterwordspacing
A.~W. Kiwelekar, P.~Patil, L.~D. Netak, and S.~U. Waikar,
  \emph{Blockchain-Based Security Services for Fog Computing}.\hskip 1em plus
  0.5em minus 0.4em\relax Cham: Springer International Publishing, 2021, pp.
  271--290. [Online]. Available:
  \url{https://doi.org/10.1007/978-3-030-57328-7_11}
\BIBentrySTDinterwordspacing

\bibitem{8314121}
M.~{Mukherjee}, L.~{Shu}, and D.~{Wang}, ``Survey of fog computing:
  Fundamental, network applications, and research challenges,'' \emph{IEEE
  Communications Surveys Tutorials}, vol.~20, no.~3, pp. 1826--1857,
  thirdquarter 2018.

\bibitem{Puliafito-Fog-2019}
\BIBentryALTinterwordspacing
C.~Puliafito, E.~Mingozzi, F.~Longo, A.~Puliafito, and O.~Rana, ``Fog computing
  for the internet of things: A survey,'' \emph{ACM Trans. Internet Technol.},
  vol.~19, no.~2, pp. 18:1--18:41, Apr. 2019. [Online]. Available:
  \url{http://doi.acm.org/10.1145/3301443}
\BIBentrySTDinterwordspacing

\bibitem{10156084}
A.~A. Baktyan and A.~T. Zahary, ``A review on cloud and fog computing
  integration for iot: Platforms perspective,'' \emph{EAI Endorsed Transactions
  on Internet of Things}, vol.~4, no.~14, 12 2018.

\bibitem{7033259}
M.~{Yannuzzi}, R.~{Milito}, R.~{Serral-Gracià}, D.~{Montero}, and
  M.~{Nemirovsky}, ``Key ingredients in an iot recipe: Fog computing, cloud
  computing, and more fog computing,'' in \emph{2014 IEEE 19th International
  Workshop on Computer Aided Modeling and Design of Communication Links and
  Networks (CAMAD)}, Dec 2014, pp. 325--329.

\bibitem{Bonomi-2342513}
\BIBentryALTinterwordspacing
F.~Bonomi, R.~Milito, J.~Zhu, and S.~Addepalli, ``Fog computing and its role in
  the internet of things,'' in \emph{Proceedings of the First Edition of the
  MCC Workshop on Mobile Cloud Computing}, ser. MCC '12.\hskip 1em plus 0.5em
  minus 0.4em\relax New York, NY, USA: ACM, 2012, pp. 13--16. [Online].
  Available: \url{http://doi.acm.org/10.1145/2342509.2342513}
\BIBentrySTDinterwordspacing

\bibitem{9139976}
T.~{Qiu}, J.~{Chi}, X.~{Zhou}, Z.~{Ning}, M.~{Atiquzzaman}, and D.~O. {Wu},
  ``Edge computing in industrial internet of things: Architecture, advances and
  challenges,'' \emph{IEEE Communications Surveys Tutorials}, pp. 1--1, 2020.

\bibitem{pal2020security}
S.~Pal, M.~Hitchens, T.~Rabehaja, and S.~Mukhopadhyay, ``Security requirements
  for the internet of things: A systematic approach,'' \emph{Sensors}, vol.~20,
  no.~20, p. 5897, 2020.

\bibitem{WANG201910}
\BIBentryALTinterwordspacing
X.~Wang, X.~Zha, W.~Ni, R.~P. Liu, Y.~J. Guo, X.~Niu, and K.~Zheng, ``Survey on
  blockchain for internet of things,'' \emph{Computer Communications}, vol.
  136, pp. 10 -- 29, 2019. [Online]. Available:
  \url{http://www.sciencedirect.com/science/article/pii/S0140366418306881}
\BIBentrySTDinterwordspacing

\bibitem{9358215}
L.~Da~Xu, Y.~Lu, and L.~Li, ``Embedding blockchain technology into iot for
  security: A survey,'' \emph{IEEE Internet of Things Journal}, pp. 1--1, 2021.

\bibitem{khan2018iot}
M.~A. Khan and K.~Salah, ``Iot security: Review, blockchain solutions, and open
  challenges,'' \emph{Future Generation Computer Systems}, vol.~82, pp.
  395--411, 2018.

\bibitem{eberhardt2017or}
J.~Eberhardt and S.~Tai, ``On or off the blockchain? insights on off-chaining
  computation and data,'' in \emph{European Conference on Service-Oriented and
  Cloud Computing}.\hskip 1em plus 0.5em minus 0.4em\relax Springer, 2017, pp.
  3--15.

\bibitem{8109357}
S.~Pal, M.~Hitchens, and V.~Varadharajan, ``Towards a secure access control
  architecture for the internet of things,'' in \emph{2017 IEEE 42nd Conference
  on Local Computer Networks (LCN)}, Oct 2017, pp. 219--222.

\bibitem{8011213}
E.~Karafiloski and A.~Mishev, ``Blockchain solutions for big data challenges: A
  literature review,'' in \emph{IEEE EUROCON 2017 -17th International
  Conference on Smart Technologies}, 2017, pp. 763--768.

\bibitem{SHARMA2020475}
\BIBentryALTinterwordspacing
A.~Sharma, E.~S. Pilli, A.~P. Mazumdar, and P.~Gera, ``Towards trustworthy
  internet of things: A survey on trust management applications and schemes,''
  \emph{Computer Communications}, vol. 160, pp. 475 -- 493, 2020. [Online].
  Available:
  \url{http://www.sciencedirect.com/science/article/pii/S0140366419319073}
\BIBentrySTDinterwordspacing

\bibitem{voas2018internet}
J.~Voas, R.~Kuhn, P.~Laplante, and S.~Applebaum, ``Internet of things (iot)
  trust concerns (draft),'' National Institute of Standards and Technology,
  Tech. Rep., 2018.

\bibitem{wu2021blockchain}
X.~Wu and J.~Liang, ``A blockchain-based trust management method for internet
  of things,'' \emph{Pervasive and Mobile Computing}, vol.~72, p. 101330, 2021.

\bibitem{DOUGLAS2020100004}
\BIBentryALTinterwordspacing
A.~Douglas, R.~Holloway, J.~Lohr, E.~Morgan, and K.~Harfoush, ``Blockchains for
  constrained edge devices,'' \emph{Blockchain: Research and Applications},
  vol.~1, no.~1, p. 100004, 2020. [Online]. Available:
  \url{https://www.sciencedirect.com/science/article/pii/S209672092030004X}
\BIBentrySTDinterwordspacing

\bibitem{101007-39393922}
V.~Gligor and J.~M. Wing, ``Towards a theory of trust in networks of humans and
  computers,'' in \emph{Security Protocols XIX}, B.~Christianson, B.~Crispo,
  J.~Malcolm, and F.~Stajano, Eds.\hskip 1em plus 0.5em minus 0.4em\relax
  Berlin, Heidelberg: Springer Berlin Heidelberg, 2011, pp. 223--242.

\bibitem{6513398}
F.~Bao, I.~Chen, and J.~Guo, ``Scalable, adaptive and survivable trust
  management for community of interest based internet of things systems,'' in
  \emph{2013 IEEE Eleventh International Symposium on Autonomous Decentralized
  Systems (ISADS)}, March 2013, pp. 1--7.

\bibitem{8328674}
D.~Ferraris, C.~Fernandez-Gago, and J.~Lopez, ``A trust-by-design framework for
  the internet of things,'' in \emph{2018 9th IFIP International Conference on
  New Technologies, Mobility and Security (NTMS)}, Feb 2018, pp. 1--4.

\bibitem{8250381}
J.~{Wang}, H.~{Wang}, H.~{Zhang}, and N.~{Cao}, ``Trust and attribute-based
  dynamic access control model for internet of things,'' in \emph{2017
  International Conference on Cyber-Enabled Distributed Computing and Knowledge
  Discovery (CyberC)}, Oct 2017, pp. 342--345.

\bibitem{Sabater2005}
\BIBentryALTinterwordspacing
J.~Sabater and C.~Sierra, ``Review on computational trust and reputation
  models,'' \emph{Artificial Intelligence Review}, vol.~24, no.~1, pp. 33--60,
  Sep 2005. [Online]. Available:
  \url{https://doi.org/10.1007/s10462-004-0041-5}
\BIBentrySTDinterwordspacing

\bibitem{3205977-3205993}
\BIBentryALTinterwordspacing
R.~Di~Pietro, X.~Salleras, M.~Signorini, and E.~Waisbard, ``A blockchain-based
  trust system for the internet of things,'' in \emph{Proceedings of the 23nd
  ACM on Symposium on Access Control Models and Technologies}, ser. SACMAT
  ’18.\hskip 1em plus 0.5em minus 0.4em\relax New York, NY, USA: Association
  for Computing Machinery, 2018, p. 77–83. [Online]. Available:
  \url{https://doi.org/10.1145/3205977.3205993}
\BIBentrySTDinterwordspacing

\bibitem{3322431-3326327}
\BIBentryALTinterwordspacing
B.~Tang, H.~Kang, J.~Fan, Q.~Li, and R.~Sandhu, ``Iot passport: A
  blockchain-based trust framework for collaborative internet-of-things,'' in
  \emph{Proceedings of the 24th ACM Symposium on Access Control Models and
  Technologies}, ser. SACMAT ’19.\hskip 1em plus 0.5em minus 0.4em\relax New
  York, NY, USA: Association for Computing Machinery, 2019, p. 83–92.
  [Online]. Available: \url{https://doi.org/10.1145/3322431.3326327}
\BIBentrySTDinterwordspacing

\bibitem{putra-9422738}
G.~D. Putra, V.~Dedeoglu, S.~S. Kanhere, R.~Jurdak, and A.~Ignjatovic,
  ``Trust-based blockchain authorization for iot,'' \emph{IEEE Transactions on
  Network and Service Management}, pp. 1--1, 2021.

\bibitem{7821686}
S.~A. {Al-Qaseemi}, H.~A. {Almulhim}, M.~F. {Almulhim}, and S.~R. {Chaudhry},
  ``Iot architecture challenges and issues: Lack of standardization,'' in
  \emph{2016 Future Technologies Conference (FTC)}, Dec 2016, pp. 731--738.

\bibitem{7785888}
E.~{Kovacs}, M.~{Bauer}, J.~{Kim}, J.~{Yun}, F.~{Le Gall}, and M.~{Zhao},
  ``Standards-based worldwide semantic interoperability for iot,'' \emph{IEEE
  Communications Magazine}, vol.~54, no.~12, pp. 40--46, December 2016.

\bibitem{Pal18}
\BIBentryALTinterwordspacing
A.~Pal, H.~K. Rath, S.~Shailendra, and A.~Bhattacharyya, ``Iot standardization:
  The road ahead,'' in \emph{Internet of Things}, J.~Sen, Ed.\hskip 1em plus
  0.5em minus 0.4em\relax Rijeka: IntechOpen, 2018, ch.~3. [Online]. Available:
  \url{https://doi.org/10.5772/intechopen.75137}
\BIBentrySTDinterwordspacing

\bibitem{101007978555}
R.~S. Malyan and A.~K. Madan, ``Blockchain technology as a tool to manage
  digital identity: A conceptual study,'' in \emph{Advances in Manufacturing
  and Industrial Engineering}, R.~M. Singari, K.~Mathiyazhagan, and H.~Kumar,
  Eds.\hskip 1em plus 0.5em minus 0.4em\relax Singapore: Springer Singapore,
  2021, pp. 635--647.

\bibitem{mahalle2010identity}
P.~Mahalle, S.~Babar, N.~R. Prasad, and R.~Prasad, ``Identity management
  framework towards internet of things (iot): Roadmap and key challenges,'' in
  \emph{International Conference on Network Security and Applications}.\hskip
  1em plus 0.5em minus 0.4em\relax Springer, 2010, pp. 430--439.

\bibitem{8726730}
A.~S. Omar and O.~Basir, ``Identity management in iot networks using blockchain
  and smart contracts,'' in \emph{2018 IEEE International Conference on
  Internet of Things (iThings) and IEEE Green Computing and Communications
  (GreenCom) and IEEE Cyber, Physical and Social Computing (CPSCom) and IEEE
  Smart Data (SmartData)}, 2018, pp. 994--1000.

\bibitem{pal2018modeling}
S.~Pal, M.~Hitchens, and V.~Varadharajan, ``Modeling identity for the internet
  of things: Survey, classification and trends,'' in \emph{2018 12th
  International Conference on Sensing Technology (ICST)}.\hskip 1em plus 0.5em
  minus 0.4em\relax IEEE, 2018, pp. 45--51.

\bibitem{1011452903150-2911710}
\BIBentryALTinterwordspacing
G.~Alp\'{a}r, L.~Batina, L.~Batten, V.~Moonsamy, A.~Krasnova, A.~Guellier, and
  I.~Natgunanathan, ``New directions in iot privacy using attribute-based
  authentication,'' in \emph{Proceedings of the ACM International Conference on
  Computing Frontiers}, ser. CF ’16.\hskip 1em plus 0.5em minus 0.4em\relax
  New York, NY, USA: Association for Computing Machinery, 2016, p. 461–466.
  [Online]. Available: \url{https://doi.org/10.1145/2903150.2911710}
\BIBentrySTDinterwordspacing

\bibitem{1011451866855-1866866}
\BIBentryALTinterwordspacing
A.~F. Gomez-Skarmeta, P.~Martinez-Julia, J.~Girao, and A.~Sarma, ``Identity
  based architecture for secure communication in future internet,'' in
  \emph{Proceedings of the 6th ACM Workshop on Digital Identity Management},
  ser. DIM ’10.\hskip 1em plus 0.5em minus 0.4em\relax New York, NY, USA:
  Association for Computing Machinery, 2010, p. 45–48. [Online]. Available:
  \url{https://doi.org/10.1145/1866855.1866866}
\BIBentrySTDinterwordspacing

\bibitem{PAL201957}
\BIBentryALTinterwordspacing
S.~Pal, M.~Hitchens, V.~Varadharajan, and T.~Rabehaja, ``Policy-based access
  control for constrained healthcare resources in the context of the internet
  of things,'' \emph{Journal of Network and Computer Applications}, vol. 139,
  pp. 57 -- 74, 2019. [Online]. Available:
  \url{http://www.sciencedirect.com/science/article/pii/S1084804519301377}
\BIBentrySTDinterwordspacing

\bibitem{zhu2018identity}
X.~Zhu and Y.~Badr, ``Identity management systems for the internet of things: a
  survey towards blockchain solutions,'' \emph{Sensors}, vol.~18, no.~12, p.
  4215, 2018.

\bibitem{8726747}
X.~{Zhu} and Y.~{Badr}, ``A survey on blockchain-based identity management
  systems for the internet of things,'' in \emph{2018 IEEE International
  Conference on Internet of Things (iThings) and IEEE Green Computing and
  Communications (GreenCom) and IEEE Cyber, Physical and Social Computing
  (CPSCom) and IEEE Smart Data (SmartData)}, 2018, pp. 1568--1573.

\bibitem{bouras2021lightweight}
M.~A. Bouras, Q.~Lu, S.~Dhelim, and H.~Ning, ``A lightweight blockchain-based
  iot identity management approach,'' \emph{Future Internet}, vol.~13, no.~2,
  p.~24, 2021.

\bibitem{9323061}
S.~Singh, A.~S. M.~S. Hosen, and B.~Yoon, ``Blockchain security attacks,
  challenges, and solutions for the future distributed iot network,''
  \emph{IEEE Access}, vol.~9, pp. 13\,938--13\,959, 2021.

\bibitem{9013321}
Y.~Nakamura, Y.~Zhang, M.~Sasabe, and S.~Kasahara, ``Capability-based access
  control for the internet of things: An ethereum blockchain-based scheme,'' in
  \emph{2019 IEEE Global Communications Conference (GLOBECOM)}, 2019, pp. 1--6.

\bibitem{8592253}
J.~Golosova and A.~Romanovs, ``The advantages and disadvantages of the
  blockchain technology,'' in \emph{2018 IEEE 6th Workshop on Advances in
  Information, Electronic and Electrical Engineering (AIEEE)}, 2018, pp. 1--6.

\bibitem{9422738}
G.~D. Putra, V.~Dedeoglu, S.~S. Kanhere, R.~Jurdak, and A.~Ignjatovic,
  ``Trust-based blockchain authorization for iot,'' \emph{IEEE Transactions on
  Network and Service Management}, pp. 1--1, 2021.

\bibitem{9382024}
S.~Hu, Y.-C. Liang, Z.~Xiong, and D.~Niyato, ``Blockchain and artificial
  intelligence for dynamic resource sharing in 6g and beyond,'' \emph{IEEE
  Wireless Communications}, pp. 1--7, 2021.

\bibitem{hussain2020machine}
F.~Hussain, R.~Hussain, S.~A. Hassan, and E.~Hossain, ``Machine learning in iot
  security: Current solutions and future challenges,'' \emph{IEEE
  Communications Surveys \& Tutorials}, vol.~22, no.~3, pp. 1686--1721, 2020.

\bibitem{salah2019blockchain}
K.~Salah, M.~H.~U. Rehman, N.~Nizamuddin, and A.~Al-Fuqaha, ``Blockchain for
  ai: Review and open research challenges,'' \emph{IEEE Access}, vol.~7, pp.
  10\,127--10\,149, 2019.

\end{thebibliography}

\end{document}